\documentclass[preprint,nofootinbib,amsmath,prd,aps,superscriptaddress,tightenlines,12pt]{revtex4}
\usepackage{graphicx}
\usepackage{graphics}
\usepackage{amsmath}
\usepackage{hyperref}
\usepackage{subfigure}

\def\lsim{\mathrel{\rlap{\lower4pt\hbox{\hskip1pt$\sim$}}
    \raise1pt\hbox{$<$}}}                
\def\gsim{\mathrel{\rlap{\lower4pt\hbox{\hskip1pt$\sim$}}
    \raise1pt\hbox{$>$}}}                

\def\OMIT#1{}

\newcommand{\be}{\begin{eqnarray}}
\newcommand{\ee}{\end{eqnarray}}

\newcommand{\nn}{\nonumber}

\newcommand{\bn}{{\bar n}}
\newcommand{\bea}{\begin{eqnarray}}
\newcommand{\eea}{\end{eqnarray}}

\def\lsim{\mathrel{\rlap{\lower4pt\hbox{\hskip1pt$\sim$}}
    \raise1pt\hbox{$<$}}}                
\def\gsim{\mathrel{\rlap{\lower4pt\hbox{\hskip1pt$\sim$}}
    \raise1pt\hbox{$>$}}}                

\def\OMIT#1{}

\begin{document}
\setlength\baselineskip{17pt}


\title{\bf  The  1-Jettiness DIS  event shape: NNLL + NLO  results}

\author{Zhong-Bo Kang}
\affiliation{Los Alamos National Laboratory,
                   Theoretical Division,
                   Los Alamos, NM 87545}
                   
\author{Xiaohui Liu}
\affiliation{High Energy Division, 
                  Argonne National Laboratory, 
                  Argonne, IL 60439}
\affiliation{Department of Physics and Astronomy, 
                   Northwestern University,
                   Evanston, IL 60208}
                   
\author{ Sonny Mantry}
\affiliation{High Energy Division, 
                  Argonne National Laboratory, 
                  Argonne, IL 60439}
\affiliation{Department of Physics and Astronomy, 
                   Northwestern University,
                   Evanston, IL 60208}




\newpage
\makebox[6.5in][r]{\hfill ANL-HEP-PR-13-18}


\begin{abstract}
  \vspace*{0.3cm}
  
We present results for the complete NNLL+NLO ($\sim \alpha_s$) 1-jettiness ($\tau_1$) event shape distribution for single jet ($J$) production in electron-nucleus ($N_A$) collisions $e^-  + N_A \to e^-  + J + X$, in the deep inelastic scattering (DIS) region where the hard scale is  set by the jet transverse momentum $P_{J_T}$.  These results cover the entire $\tau_1$-spectrum including the resummation  ($\tau_1\ll P_{J_T}$) and fixed-order  ($\tau_1\sim P_{J_T}$) perturbative QCD  regions. They incorporate non-perturbative soft radiation effects,  the anti-$k_T$ jet algorithm in the fixed-order calculation, and  a smooth matching between the resummation and fixed-order perturbative QCD regions. The matching smoothly connects the spectrum in the resummation region, which can be computed without reference to an external jet algorithm, and the fixed-order region where an explicit jet algorithm must be specified. Our code, used for generating the numerical results, is flexible enough to incorporate different jet algorithms for the fixed-order calculation. We also perform a jet shape analysis, defined within the 1-jettiness framework, which allows one to control the amount of radiation included in the definition of the final state jet. This formalism can allow for detailed studies of jet energy-loss mechanisms and nuclear medium effects. The analysis presented here can be used for precision studies of QCD and as a probe of nuclear dynamics using data collected at HERA and in proposed future electron-ion colliders
 such as the EIC and the LHeC.

\end{abstract}

\maketitle

\newpage
\tableofcontents

\section{Introduction}

Event shapes analyses are powerful probes of QCD dynamics and have now been applied for a variety of processes. Event shapes for DIS processes   were first introduced and developed in Refs.~\cite{Antonelli:1999kx,Dasgupta:2001sh,Dasgupta:2001eq,Dasgupta:2002bw}. Thrust~\cite{Antonelli:1999kx} and Broadening~\cite{Dasgupta:2001eq} distributions were studied at the next-to-leading logarithmic (NLL) level of accuracy and matched at ${\cal O}(\alpha_s)$ to  fixed-order results. A numerical comparison was also done against ${\cal O}(\alpha_s^2)$ results~\cite{Catani:1996vz,Graudenz:1997gv}. Thrust distributions have also been measured at HERA by the H1~\cite{Adloff:1997gq,Aktas:2005tz,Adloff:1999gn} and ZEUS~\cite{Breitweg:1997ug,Chekanov:2002xk,Chekanov:2006hv} collaborations.

A new event shape called N-jettiness ($\tau_N$) \cite{Stewart:2009yx,Stewart:2010tn} was recently introduced as tool to inclusively veto jets at the LHC. It quantifies the amount and shape of the radiation in the final state for events with $N$ jets. This makes it ideal for an analyses of exclusive jet production at hadron colliders. The veto on additional jets is applied by going to the limit $\tau_N\to 0$. In this limit, energetic radiation is allowed only along the $N$ jet and two beam directions. Any radiation at wide angles from the jet and beam directions is restricted to be soft with energy $E\sim \tau_N$; effectively acting as a veto on additional hard radiation or jets. Since the limit $\tau_N\to 0$  is dominated by energetic radiation that is collinear with one of the $N$ jet directions or one of the beam directions and soft radiation ($E\sim \tau_N$) everywhere else, it can be treated using the Soft Collinear Effective Theory (SCET) \cite{Bauer:2000ew,Bauer:2000yr,Bauer:2001ct,Bauer:2001yt,Bauer:2002nz,Beneke:2002ph}, which facilitates resummation of the associated large Sudakov logarithms. For LHC processes, numerical results have now been obtained in the resummation region for beam thrust ($0$-jettiness) distributions for Drell-Yan processes \cite{Stewart:2009yx,Stewart:2010pd} and Higgs production \cite{Berger:2010xi},  threshold resummation in gauge boson production with two final-state jets \cite{Liu:2012zg}, and the jet mass spectrum for Higgs production with one final-state jet~\cite{Jouttenus:2013hs}. 

Recently~\cite{Kang:2012zr}, the 1-jettiness ($\tau_1$) event shape was proposed for single jet ($J$) production in the DIS process
\bea
\label{process}
e^- + N_A \to J + X,
\eea
where $N_A$ denotes a nucleus with atomic weight $A$.  In particular, a factorization and resummation framework was proposed and derived for the observable 
\bea
\label{obs}
d\sigma_A \equiv \frac{d^3\sigma (e^- + N_A\to J + X)}{dy\> dP_{J_T}\>d\tau_1},
\eea
in the limit
\bea
\label{pscond}
\tau_1 \ll P_{J_T},
\eea
where $P_{J_T}$ and $y$ denote the transverse momentum and rapidity of the final state jet. 
Numerical results were also presented with a resummation of  Sudakov logarithms of the form $\alpha_s ^n \ln^{2m} (\tau_1/P_{J_T})$ for $m \leq n$, at the next-to-leading logarithmic (NLL) level of accuracy for a proton $(A=1)$ target. In Ref.~\cite{Kang:2013wca}, the results were extended to include resummation at the next-to-next-leading logarithmic (NNLL) level of accuracy. Furthermore, numerical results were presented for a wide range of nuclear targets: proton, Carbon, Calcium, Iron, Gold, and Uranium. Shortly thereafter, NNLL resummation results for a proton target  were also presented in Ref.~\cite{Kang:2013nha}. In addition,  Ref.~\cite{Kang:2013nha} introduced two new definitions of 1-jettiness, with the corresponding factorization formulae, associated with different choices of reference vectors used to define the 1-jettiness event shape. We note that the 1-jettiness event shape $\tau_1$ considered here, is distinct from those considered in the previous works of Refs.~\cite{Antonelli:1999kx,Dasgupta:2001sh,Dasgupta:2001eq,Dasgupta:2002bw}.  For more details on the differences between the different types of DIS event shapes, we refer the reader to section III-B of Ref.~\cite{Kang:2013nha} which uses the notation $\tau_1^a$ for the DIS 1-jettiness event shape $\tau_1$ considered here and first introduced in Refs.~\cite{Kang:2012zr,Kang:2013wca}.

So far all previous works on 1-jettiness for DIS have been focused on the resummation region, defined by the region in Eq.~(\ref{pscond}), where the Sudakov logarithms of $\tau_1/P_{J_T}$ are large and the cross-section is dominated by terms singular in the $\tau_1\to 0$ limit. In order to obtain the full spectrum, results are needed in the region of large 1-jettiness
\bea
\label{largetau1}
\tau_1 \sim P_{J_T},
\eea
where the non-singular terms become important and the perturbative QCD framework is appropriate. In addition, one must match the resummation and fixed-order regions, Eqs.~(\ref{pscond}) and (\ref{largetau1}) respectively, in order to have a smooth and continuous spectrum for all values of $\tau_1$. 

The focus of this paper is to present numerical results for the next-to-leading order (NLO) contribution in the fixed-order region, defined as the order $\alpha_s$ contribution, and its matching to the resummation region; i.e. the full NNLL + NLO($\sim \alpha_s$) 1-jettiness spectrum. 
\begin{figure}
\subfigure  {\label{fig:subfig1}\includegraphics[scale=0.2]{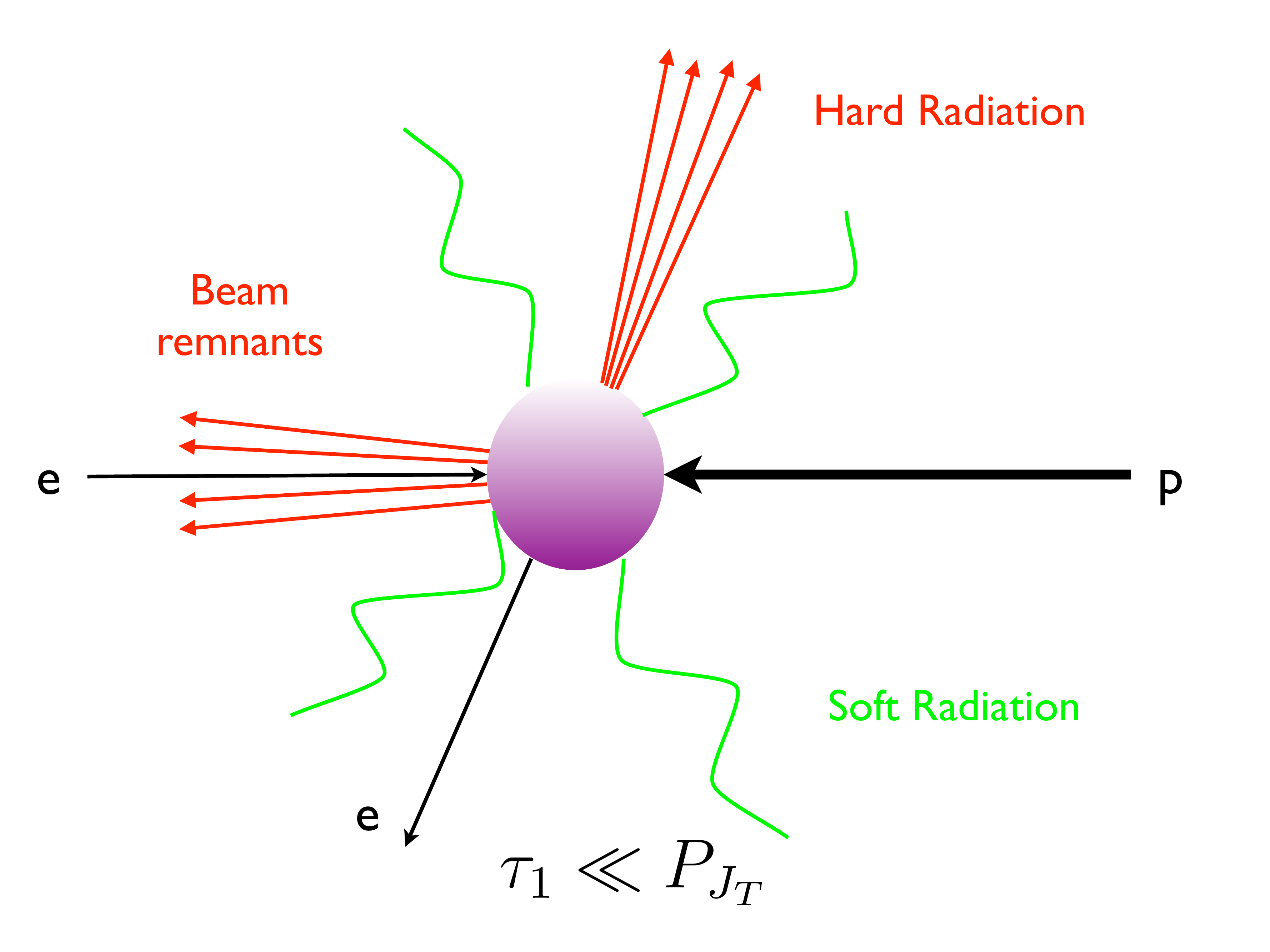}}
\subfigure {\label{fig:subfig2}\includegraphics[scale=0.2]{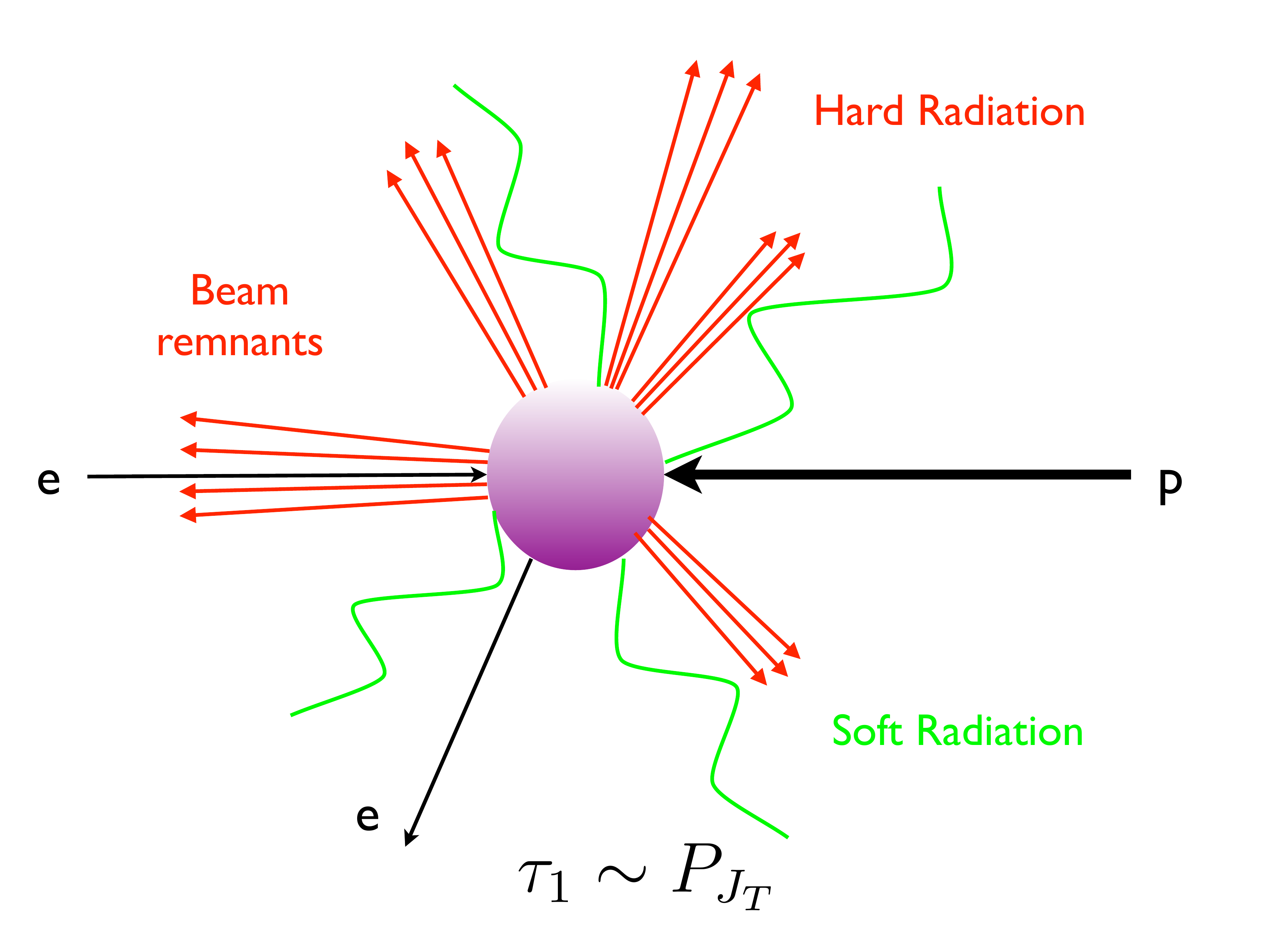}}
\caption{Schematic figure of the process $e^- + p \to J + X$ in the limit $\tau_1\ll P_{J_T}$. The restriction $\tau_1\ll P_{J_T}$ (left panel) allows only soft radiation between the beam and jet directions. In the region of large 1-jettiness $\tau_1 \sim P_{J_T}$ (right panel), additional hard radiation is allowed at wide angles from the leading jet and beam directions.}
\label{fig:process}
\end{figure}
In the resummation region $\tau_1 \ll P_{J_T}$, the dependence on the jet algorithm is power suppressed~\cite{Stewart:2010tn} in $\tau_1/P_{J_T}$. This property was exploited in Refs.~\cite{Kang:2012zr,Kang:2013wca,Kang:2013nha} to achieve resummation without having to implement an explicit jet finding algorithm. However, in the fixed-order region, $\tau_1 \sim P_{J_T}$ so that the dependence on the jet algorithm is no longer power suppressed. This situation is schematically illustrated in Fig.~\ref{fig:process}. The left and right panels show typical configurations in the resummation and fixed-order regions respectively. Since the resummation region corresponds to configurations with one narrow jet and only soft radiation between the jet and beam directions, different jet algorithms will yield the same energy and direction for the leading jet, so that one obtains the same reference jet axis used in the calculation of 1-jettiness; up to power corrections in $\tau_1/P_{J_T}$. On the other hand, in the fixed-order region $\tau_1\sim P_{J_T}$, hard  radiation is allowed in multiple directions at wide angles from each other. In this case, different jet algorithms will yield a different energy and direction for the leading jet,
which can have order one effects on the 1-jettiness distribution.  Thus, for the NNLL+NLO 1-jettiness spectrum, different jet algorithms could lead to significant differences in the  fixed-order region $\tau_1 \sim P_{J_T}$, but the distributions are expected to converge in the limit $\tau_1 \ll P_{J_T}$. 

In this work, we extend the results of Refs.~\cite{Kang:2012zr,Kang:2013wca} to give the full NNLL+NLO $\tau_1$-spectrum. We explicitly incorporate the anti-$k_T$ jet algorithm~\cite{Cacciari:2008gp} with jet radius $R=1.0$ to find the leading jet, and correspondingly the reference jet axis used in the calculation of $\tau_1$, in the NLO calculation in the fixed-order region $\tau_1\sim P_{J_T}$. However, our code used for the numerical analysis is flexible enough to easily adapt different types of jet algorithms and different values for the jet radius. We perform several consistency checks including numerically demonstrating the cancellation of all infrared divergences occurring  in the NLO calculation,   the convergence of the NLO result with the  NNLL resummed result expanded to fixed-order in the singular limit $\tau_1\to 0$, and the convergence of the NNLL+NLO $\tau_1$-distribution to the fixed-order NLO result in the perturbative QCD region of large $\tau_1$. We also incorporate non-perturbative soft radiation effects in the region $\tau_1\sim \Lambda_{QCD}$ through a phenomenological model for the soft function such that it correctly reproduces the soft function scale dependence and reduces to the perturbative result for $\tau_1\gg \Lambda_{QCD}$. Finally, we perform a jet-shape analysis, defined within the 1-jettiness framework, which allows one to change the  amount of radiation included in the final state jet definition. Such an analysis can be a powerful probe of energy loss mechanisms in final state jets and nuclear medium effects on the propagation of hard partons.

As discussed in Refs.~\cite{Kang:2012zr,Kang:2013wca}, the formalism and results presented here can be applied to data collected at HERA and in proposed electron-ion colliders~\cite{Boer:2011fh, AbelleiraFernandez:2012cc,AbelleiraFernandez:2012ni}, for precision studies of QCD and as a probe of nuclear dynamics. In this work, we give numerical results only for the proton target but similar results can be easily derived for heavier nuclei, as was done for NNLL resummation spectrum in Ref.~\cite{Kang:2013wca}. We leave such an analysis for future work.


\section{Formalism}

For a more detailed review of the formalism of 1-jettiness applied to DIS processes, we refer the reader to Refs.~\cite{Kang:2013wca,Kang:2013ata}. Here we give only a brief overview and establish relevant notation and definitions. 

\subsection{Kinematics}
We work in the center of mass frame of the electron and the \textit{average} nucleon momentum in the nucleus. In this frame, the electron and nucleus momenta $p_e$ and $p_A$ are defined as
\bea
p_e^\mu &=& (p_e^0,\vec{p_e}), \qquad P_A^\mu = A (p_e^0,-\vec{p}_e),
\eea
where $A$ is the atomic weight of the nucleus. We neglect the electron mass for simplicity and write
\bea
p_e^0=|\vec{p}_e| = \frac{Q_e}{2},
\eea
so that hadronic center of mass energy is given by
\bea
s = (p_e + P_A)^2 =A\> Q_e^2.
\eea
In terms of the light-like vectors $n_A=(1,0,0,1)$ and $\bn_A=(1,0,0,-1)$, the electron and nucleus momentum can be written as
\bea
P_A^\mu = A\frac{Q_e}{2} n_A^\mu, \qquad p_e^\mu = \frac{Q_e}{2}\bn_A^\mu.
\eea

\subsection{1-Jettiness and Jet Algorithms}
\label{jetalgo}
The 1-jettiness global event shape $\tau_1$ is defined as
\bea
\label{1-jettiness}
\tau_1 &=& \sum_k \text{min} \Big \{ \frac{2q_A\cdot p_k}{Q_a}, \frac{2q_J\cdot p_k }{Q_J}\Big \},
\eea
where the sum is over all final state particles (except the final state lepton) with momenta $p_k$. The light-like four-vectors $q_A$ and $q_J$ denote reference vectors along the nuclear beam and final state jet directions respectively. The constants $Q_a$ and $Q_J$ are of the order of the hard scale and their choices are not unique; different choices can be interpreted as different definitions of $\tau_1$.

The beam reference vector is generally chosen to align with the z-axis and we make the specific choice
\bea
\label{qAdef}
q_A = x_A P_A.
\eea
In general, the choice of the jet reference vector $q_J$ depends on the jet algorithm used. Procedurally, a standard jet algorithm~\cite{Cacciari:2011ma} such as  $k_T$, anti-$k_T$, or Cambridge-Aachen is used to find all jets in a given event. The momentum of the leading jet, denoted as $K_J$, has the general form
\bea
\label{Kjet}
K_J = (E_K \cosh y_K, \vec{K}_{J_T},  E_K \sinh y_K), \qquad E_K^2 = K_{J_T}^2 +M_J^2,
\eea
where $M_J$ denotes the jet mass. The massless jet reference vector $q_J$ can be constructed from the leading jet momentum in terms of its transverse momentum and rapidity as
\bea
\label{qJdef}
q_J=(K_{J_T}\cosh y_K, \vec{K}_{J_T},K_{J_T}\sinh y_K).
\eea
Since, in general, different jet algorithms will yield a different  leading jet, the jet reference vector $q_J$ depends on the jet algorithm used. Note that the reference vector $q_J$  is distinct  from the \textit{total} leading jet momentum $K_J$. In particular, $q_J$ is defined to be a massless vector constructed out of the transverse momentum and rapidity of the leading jet. On the other hand, in general, the leading jet will have a  non-zero mass $M_J^2=K_J^2\neq 0$. We now choose the $Q_a$ and $Q_J$ constants in Eq.~(\ref{1-jettiness}) as
\bea
\label{QaJ}
 Q_a=x_A A Q_e, \qquad  Q_J = 2K_{J_T}\cosh y_K,
\eea
where $x_A$ is the nucleus momentum fraction carried by the initial parton in the hard interaction. 

Note that  the jet algorithm has only been used to determine the jet reference vector $q_J$. The ``1-jettiness jet momentum" $P_J$, whose transverse momentum and rapidity are measured in Eq.(\ref{obs}), has not yet been defined and is in general distinct from the leading jet momentum $K_J$ found by the external jet  algorithm. In particular, the 1-jettiness momentum $P_J$ is defined  as
\bea
\label{pjet}
P_J &=& \sum_k p_k \>\theta (\frac{2q_A\cdot p_k}{Q_a} - \frac{2q_J\cdot p_k}{Q_J} ),
\eea
where $q_A$ is the beam reference vector defined in Eq.~(\ref{qAdef}) and the jet reference vector $q_J$ is determined by the external jet algorithm as in Eq.~(\ref{qJdef}). The transverse momentum $P_{J_T}$ and the rapidity $y$ appearing in Eq.~(\ref{obs}), correspond to those of the 1-jettiness jet momentum defined in Eq.~(\ref{pjet}).

In the resummation region $\tau_1 \ll P_{J_T}$,  corresponding to configurations that look like the left panel of Fig.~\ref{fig:process}, different jet algorithms will yield the same reference vector $q_J$ up to power corrections. i.e. the resummation region corresponds to a single hard jet well separated from the beam direction, with only soft radiation between the beam and jet directions. In this case, differences between jet algorithms correspond to differences in the amount of soft radiation included in the jet. The soft radiation only affects the mass of the jet; not its energy and direction. In particular, the transverse momentum $K_{J_T}$ and rapidity $y_K$ of the leading jet are insensitive to soft radiation, up to power corrections in $\tau_1/P_{J_T}$. Consequently, the jet reference vector $q_J$, given in Eq.(\ref{qJdef}), is independent of the jet algorithm in the resummation region. The 1-jettiness momentum $P_J$, determined in terms of $q_J$ as in Eq.(\ref{pjet}), is then also independent of the jet algorithm in this region. In fact, in this region (see left panel of Fig.~\ref{fig:process}) one will find that $P_{J_T}= K_{J_T}$ and $y= y_K$, up to power corrections in $\tau_1/P_{J_T}$. Thus, in the resummation region, the jet reference vector $q_J$ and the constant $Q_J$ can now be written as
\bea
\label{qJresum}
q_J \Big |_{\tau_1 \ll P_{J_T}} \simeq (P_{J_T}\cosh y, \vec{P}_{J_T},P_{J_T}\sinh y), \qquad  Q_J\Big |_{\tau_1 \ll P_{J_T}} \simeq 2P_{J_T}\cosh y
\eea
Thus, in the limit $\tau_1 \ll P_{J_T}$, the 1-jettiness event shape and the observable in Eq.(\ref{obs}) can be computed without the explicit use of any jet algorithm. In particular, for each a priori specified values of $P_{J_T}, y$ on the LHS of Eq.(\ref{obs}), the reference vector $q_J$ is defined as in Eq.(\ref{qJresum}). Using this definition for $q_J$ in Eq.(\ref{pjet}), in the resummation region the obtained values for transverse momentum and rapidity will coincide with the a priori specified values $P_{J_T},y$; up to power corrections in $\tau_1/P_{J_T}$. Once again note that in general, $q_J$ differs from the \textit{total} 1-jettiness jet momentum $P_J$ in that it is defined to be massless and only depends on the transverse momentum and rapidity of $P_J$. On the other hand, the $P_J$ in Eq.(\ref{pjet}) will in general have a non-zero mass $P_J^2\neq 0$.

To summarize, for a priori specified values of $P_{J_T}$ and $y$ on the LHS of Eq.(\ref{obs}), together with the definitions of $q_J$, $q_A$, and $\tau_1$ in Eqs. (\ref{qJresum}), (\ref{qAdef}), and (\ref{1-jettiness}) respectively, the observable in Eq.(\ref{obs}) can be unambiguously computed without reference to an explicit jet algorithm in the resummation region $\tau_1\ll P_{J_T}$. This allowed the earlier works in Refs.~\cite{Kang:2012zr,Kang:2013wca,Kang:2013nha} to give results in the resummation region without making use of an explicit jet algorithm. 

In the fixed-order region $\tau_1 \sim P_{J_T}$, energetic radiation is allowed at wide angles from the beam and leading jet directions. This situation is illustrated in the right panel of Fig.~\ref{fig:process}. In this case, different jet algorithms will cluster hard radiation into different jets so that the leading jet momentum $K_J$ can vary significantly with the jet algorithm used. Correspondingly, the jet reference vector $q_J$ in Eq.(\ref{qJdef}) will also vary with the jet algorithm; in particular the transverse momentum $K_{J_T}$ and rapidity $y_K$ of the leading jet can vary with the jet algorithm used. As a result, 1-jettiness jet momentum $P_J$, which depends on $q_J$ as defined in Eq.(\ref{pjet}), will also vary with the jet algorithm used. Thus, unlike the resummation region, in the fixed-order region one must specify the explicit jet algorithm used in order to interpret results sensibly.

For the NLO calculation and in our numerical results, we use the anti-$k_T$~\cite{Cacciari:2008gp} jet algorithm where the distance metrics are defined as
\bea
\rho_{ij} = {\rm min}(p_{T,i}^{-1},p_{T,j}^{-1})\> \frac{\Delta R_{ij}}{R}, \qquad \rho_i = p_{T,i}^{-1},
\eea 
where $\Delta R_{ij}^2=\Delta \eta_{ij}^2+\Delta \phi_{ij}^2$ and $R$ is the jet radius parameter.
However, our numerical code is flexible enough to easily accommodate other jet algorithms. We remind the reader that such explicit  jet algorithms are only used for the purposes of defining the jet reference vector $q_J$ in terms of the leading jet momentum, as in Eq.(\ref{qJdef}). The final state jet momentum is then defined through Eq.(\ref{pjet}) and corresponds to the 1-jettiness definition of the final state jet momentum, which differs from the leading jet momentum obtained through an explicit jet algorithm. Thus, the 1-jettiness jet momentum $P_J$ depends on an explicit jet algorithm only indirectly through its dependence on the jet reference vector $q_J$, appearing in Eq.~(\ref{pjet}).

\subsection{Matching the Resummation and fixed-order Regions}

In order to obtain a continous 1-jettiness spectrum for all values of $\tau_1$, the resummation and fixed-order regions, $\tau_1 \ll P_{J_T}$ and $\tau_1 \sim P_{J_T}$ respectively, must be smoothly matched. Based on the discussion in the previous sections, this means that the 
resummation of Sudakov logarithms in $\tau_1/P_{J_T}$ must turn off in the fixed-order region $\tau_1 \sim P_{J_T}$ and the
distributions calculated using different jet algorithms must converge in the resummation region $\tau_1 \ll P_{J_T}$.  The full 1-jettiness spectrum with a matching of the resummation and fixed-order regions is given by the standard schematic formula
\bea
\label{match}
d\sigma = \big [ d\sigma_{\text{resum}} - d\sigma_{\text{resum}}^{FO} \big ] + d\sigma^{FO},
\eea
where $d\sigma_{\text{resum}}$ denotes the resummed cross section computed in the  region $\tau_1 \ll P_{J_T}$,  $d\sigma_{\text{resum}}^{FO}$ denotes this resummed cross section expanded to fixed-order perturbation theory, and $d\sigma^{FO}$ denotes the full cross section  computed to the same order in perturbative QCD. The $d\sigma^{FO}$ differs from $d\sigma_{\text{resum}}^{FO}$ by terms that are non-singular in the limit $\tau_1\to 0$. As required, in the resummation region $\tau_1 \ll P_{J_T}$, $d\sigma$ in Eq.(\ref{match}) is dominated by $d\sigma_{\text{resum}}$ due to a cancellation  between $d\sigma_{\text{resum}}^{FO}$ and $d\sigma^{FO}$, up to suppressed non-singular terms. Similarly, in the fixed-order region $\tau_1 \sim P_{J_T}$, $d\sigma$ is dominated by $d\sigma^{FO}$   due to a cancellation  between $d\sigma_{\text{resum}}$ and $d\sigma_{\text{resum}}^{FO}$, up to terms suppressed in perturbation theory.

Furthermore, as discussed in the previous sections, the explicit jet algorithm dependence is contained entirely in $d\sigma^{FO}$. On the other hand, the $d\sigma_{\text{resum}}$  and $d\sigma_{\text{resum}}^{FO}$ contributions are computed with no reference to an explicit jet algorithm; i.e. the jet reference vector $q_J$, needed to compute $\tau_1$, is given by Eq.(\ref{qJresum}) in the resummation region. As a result, in the resummation region where $d\sigma_{\text{resum}}$ dominates, distributions calculated with different jet algorithms are expected to converge. Correspondingly, in the fixed-order region where $d\sigma^{FO}$ dominates, distributions calculated with different jet algorithms could yield significant differences. 

Thus, $d\sigma$, as defined in Eq.(\ref{match}), gives the full 1-jettiness spectrum while encoding the important features of the resummation and fixed-order regions and providing a smooth matching between these two regions. We now discuss each of the terms that appear in Eq.(\ref{match}).

In the resummation region (left panel of Fig.~\ref{fig:process}), the factorization formula derived in Refs.~\cite{Kang:2012zr,Kang:2013wca} for $d\sigma_{\text{resum}}$ for the process in Eq.(\ref{process}) has the schematic form
\bea
\label{schem-1}
d\sigma_{\text{resum}} \equiv \frac{d^3\sigma_{\text{resum}}}{dy dP_{J_T} d\tau_1} &\sim &H \otimes B \otimes J \otimes {\cal S},  
\eea
where $H$ denotes the hard function that describes the physics of the hard scattering, the beam function  $B$~\cite{Fleming:2006cd,Stewart:2009yx}   describes the dynamics of the initial state PDF and the perturbative initial state radiation collinear with the beam direction,  the jet function $J$ describes the dynamics of the collinear radiation in the final state jet,  and the soft function ${\cal S}$ describes the dynamics of soft radiation ($E\sim \tau_1$) throughout the event. The beam function is matched onto the standard PDF 
\bea
\label{schem-2}
B \sim {\cal I} \otimes f,
\eea
where ${\cal I}$ is a perturbatively calculable coefficient that isolates the dynamics of perturbative initial state collinear radiation close to the beam direction.  Each of these functions are associated with a natural scale: the hard, jet, beam, and soft scales  
\bea
\label{scales}
\mu_H \sim P_{J_T},\qquad \mu_J\sim \mu_B\sim \sqrt{\tau_1 P_{J_T}}, \qquad \mu_S \sim \tau_1,
\eea
 respectively. All objects in Eqs.~(\ref{schem-1}) and (\ref{schem-2}) are evaluated at a common scale $\mu$ using renormalization group equations for each of these objects in the SCET.  

We refer the reader to Ref.~\cite{Kang:2013wca} for a detailed version of the factorization formula in Eqs. (\ref{schem-1}) and (\ref{schem-2}), including steps in  the derivation of the factorization formula and field-theoretic definitions of all the relevant objects.

The corresponding contribution $d\sigma_{\text{resum}}^{FO}$ in Eq.(\ref{match}), is obtained by setting all scales in the resummation formula equal to each other so that
\bea
\label{resumoff}
d\sigma_{\text{resum}}^{FO} &=&d\sigma_{\text{resum}} (\mu=\mu_H=\mu_J=\mu_B=\mu_S),
\eea
thereby turning off resummation and only leaving the contributions of the fixed-order SCET matrix elements that appear in Eqs.(\ref{schem-1}) and (\ref{schem-2}). 

In the fixed-order region (right panel of Fig.~\ref{fig:process}), the schematic form of the fixed-order contribution $d\sigma^{FO}$ in Eq.(\ref{match}) is given by
\bea
\label{dsFO}
d\sigma^{FO} &\sim & \int dPS \>\>\hat{{\cal F}}_{\text{meas.}}([PS])\> \big |{\cal M}\big |^2 \otimes f , 
\eea
where $dPS$ denotes the measure of integration over the final state phase space,  $|{\cal M}|^2$ denotes the UV renormalized amplitude squared for the partonic process, $f$ denotes the initial state PDF, and $ \hat{{\cal F}}_{\text{meas.}}$ denotes the measurement function that imposes the specified restrictions on the final state. In particular, for the observable in Eq.(\ref{obs}) it restricts the final state jet to have a transverse momentum and rapidity of $P_{J_T}$ and $y$ respectively and the final state radiation to have the value $\tau_1$ for the 1-jettiness event shape. In this work, these final state restrictions  are implemented numerically. In particular, we use Vegas~\cite{Hahn:2004fe} to generate phase space points and then perform numerical integrations after imposing the final state restrictions.  For each phase space point, a jet algorithm is implemented to cluster final state particles and find the leading jet. The transverse momentum ($K_{J_T}$) and rapidity ($y_K$) of the leading jet for each phase space point is then used to construct the jet reference vector $q_J$, as in Eq.(\ref{qJdef}). Along with $q_A$ given by Eq.(\ref{qAdef}) and $Q_{a,J}$ given by Eq.(\ref{QaJ}), the set of values $\tau_1,P_{J_T}, y$ is returned for each phase space point. Numerical integrations are then performed by restricting the phase space to be within specified bin sizes around specified central values for $\tau_1,P_{J_T},$ and $y$.

\section{NLO Calculation}
\label{nlocalc}

In this section we outline the procedure used for the NLO calculation ($\sim \alpha_s$) of $d\sigma^{FO}$, appearing in Eq.(\ref{match}), for the process in Eq.(\ref{process}). At LO the partonic channels are
\bea
\label{LO}
&&e^- + q_i \to e^- + q_ i,\nn \\
&&e^- + \bar{q}_i \to e^- + \bar{q}_ i
\eea
where the index $i$ runs over the  quark and antiquark flavors. The NLO contribution is given by three types of partonic channels with the real emission of an extra parton in the final state
\bea
\label{NLO}
&&e^- + q_i \to e^- + q_ i + g , \nn \\
&& e^- + \bar{q}_i \to e^- + \bar{q}_ i + g \nn \\
&&e^- + g \to e^- + q_ i + \bar{q}_i,
\eea
and the virtual corrections to the leading order channels in Eq.(\ref{LO}). 

Infrared (IR) singularities arise in these NLO calculations from the real emission of an extra parton in the final state as well as from the virtual corrections to the leading order process. In order to numerically evaluate $d\sigma^{FO}$ in Eq.(\ref{dsFO}), it becomes necessary to analytically isolate these  IR singularities. We use dimensional regularization by working in $d=4-2\epsilon$ dimensions to isolate the IR divergences as poles in $\epsilon$.  

For the virtual corrections to the leading order process in Eq.(\ref{LO}), the IR poles in $\epsilon$ can be easily extracted through analytic calculation of the virtual one loop diagrams. The channels in Eq.(\ref{NLO}), with the emission of an extra parton in the final state, give rise to soft and collinear divergences in the phase space integration when one  or a pair of partons becomes unresolved, respectively. For example, in the first two channels in Eq.(\ref{NLO}), IR singularities arise when the final state gluon becomes soft or collinear with either the initial or final state quark/antiquark. In the third channel in Eq.(\ref{NLO}), IR singularities arise when the final state quark or the anti-quark becomes collinear with the initial state gluon.

In order to implement the numerical computation of the phase space integrations for the channels in Eq.(\ref{NLO}), incorporating all the final state restrictions, it becomes necessary to first isolate the IR divergences analytically. We follow the procedure of isolating IR singularities through an appropriate parameterization of the phase space and expanding in plus-distributions. The basic idea is to map the phase space integration variables to a new set of variables $x_i$, with a range of integration $x_i \in [0,1]$. In this parameterization, the IR singularities arise when a subset of the variables $x_i$ approach zero and can be extracted as poles in $\epsilon$ by working in $d=4-2\epsilon$ dimensions. This subset of variables corresponds to the rescaled energies of unresolved partons and the relative angles between two unresolved partons, corresponding to the soft and collinear singularities respectively. 

However, different partons become unresolved in different regions of phase space, corresponding to different subsets of the $x_i$ parameterizing the IR singularities.  At NLO it was shown~\cite{Frixione:1995ms} that this can be dealt with using sector decomposition and was later extended to NNLO calculations~\cite{Czakon:2010td,Boughezal:2011jf,Boughezal:2013uia}. i.e. the phase space regions are decomposed into separate sectors such that in each sector  only single parton or a single pair of partons becomes unresolved. For instance, for $e + q_i\to e + q_i + g$ a sector decomposition is needed to isolate the cases where $g$ is parallel to the initial state $q_i$ 
or the final state $q_i$. Similarly, a sector decomposition is needed for  $e+g\to e +q_i +{\bar q}_i$  to isolate the cases where the final state $q_i$ or $\bar{q}_i$ is collinear with the initial state $g$.

Using the sector decomposition technique, the phase integrations in \textit{each sector} for the channels in Eq.(\ref{NLO}), can be brought into the schematic form
\bea
\int dPS \>\tilde{F} = \int_0^1 \prod_i\> \big [dx_i \big ]\>  x_1^{-1-a_1 \epsilon}x_2^{-1-a_2 \epsilon}\> \>F(\{x_i\}),
\eea
where on the LHS, $\tilde{F}$ schematically denotes the integrand of phase space integration and on the RHS $F(\{x_i\})$ is defined such that it is finite in the limit of any $x_i\to 0$. In this parameterization, all soft and collinear IR singularities correspond to the limits $x_1\to 0$ or $x_2\to 0$. These singularities are extracted as poles in $\epsilon$ using the standard identity
\bea
x^{-1-a\epsilon} = - \frac{1}{a \epsilon}\delta(x) + \sum_{n=0}^\infty \frac{(-\epsilon a)^n}{n!} \,
\left(\frac{\log^n x}{x} \right)_+\,,
\eea
to expand in delta-distributions and plus-distributions.
In the limit $\epsilon\to 0$, these phase space integrations can then be brought into the schematic form
\bea
\label{NLO-scheme}
\int dPS \>\tilde{F} =
\frac{A}{\epsilon^2} + \frac{B}{\epsilon} + C \,,
\eea
where the coefficients $A$, $B$, and $C$ can be now be numerically evaluated.

It is well-known that for infrared-safe observables, the IR poles must cancel between the virtual corrections to the leading order process in Eq.(\ref{LO}) and those arising from the emission of an extra parton in the processes in Eq.(\ref{NLO}), up to those that are absorbed into the PDF. This allows for an important cross-check of the phase space integrations in Eq.(\ref{NLO-scheme}). In particular, the pole terms in Eq.(\ref{NLO-scheme}), when summed over all phase space sectors in the sector decomposition technique, must accordingly cancel against the pole terms arising from the  virtual corrections to the corresponding leading order process in Eq.(\ref{LO}). We provide an explicit numerical check of this consistency condition in the numerical results section. 

Using the sector decomposition technique, we provide numerical results for $d\sigma^{FO}$ appearing in Eq.(\ref{dsFO}) and match it to the resummation region to give the full spectrum, as defined in Eq.(\ref{match}). For more details on the NLO calculation and its numerical implementation, we refer the reader to appendices~\ref{secNLO} and \ref{psdetail}.

\section{Numerical Results}

In this section we present numerical results for the NLO ($\sim \alpha_s $) calculation in perturbative QCD for the observable in Eq.(\ref{obs}), and its matching to the NNLL resummed result as given in Eq.(\ref{match}). For all results below,  for the NLO contribution $d\sigma^{FO}$ in Eq.(\ref{match}), we use the anti-$k_T$ jet algorithm~\cite{Cacciari:2008gp} with jet radius $R=1.0$ in order to determine the jet reference vector $q_J$, as described in section \ref{jetalgo}. However, we note that our code is flexible enough to easily incorporate other types of jet algorithms and vary the jet radius.

\subsection{Consistency Checks}
Before we present results for the full $\tau_1$-spectrum, we perform several consistency checks. First, we demonstrate the cancellation of IR singularities between the NLO virtual corrections to the LO processes in Eq.(\ref{LO}) and the NLO real emission contributions in Eq.(\ref{NLO}), up to collinear divergences that are absorbed into the PDF. In particular, this cancellation of IR singularities implies the condition 
\bea
\label{IRpole1}
&&\big [d\hat{\sigma}^{V}(e^-+q\to e^-+q) +d\hat{\sigma}^R(e^-+q\to e^-+q +g )\nn \\
&&-\frac{P_{qq}}{\epsilon}\otimes d\hat{\sigma}^{\rm Born}(e^-+q\to e^-+q)\big ]\Big |_{\rm IR} =0, \nn \\
\eea
for the electron-quark channel and
\bea
\label{IRpole2}
&&\big [d\hat{\sigma}^R(e^-+g\to e^-+q+\bar{q}) -\frac{P_{qg}}{\epsilon}\otimes d\hat{\sigma}^{\rm Born}(e^-+q\to e^-+q)\nn \\
&&-\frac{P_{\bar{q}g}}{\epsilon}\otimes d\hat{\sigma}^{\rm Born}(e^-+\bar{q}\to e^-+\bar{q})\big ]\Big |_{\text{IR pole terms}}=0\, ,
\eea
for the electron-gluon channel. i.e. in the electron-quark channel the sum of the real (R) and virtual (V) corrections must leave only a collinear divergence that can be absorbed by the quark PDF, corresponding to the term with the splitting function $P_{qq}$. Similarly, in the electron-gluon channel there are  only collinear divergences that can be absorbed by the gluon PDF, corresponding to the $P_{qg}$ and $P_{\bar{q}g}$ splitting function terms.
\begin{figure}[h!]
\begin{minipage}[b]{3.2in}
  \includegraphics[width=3.2in,angle=0]{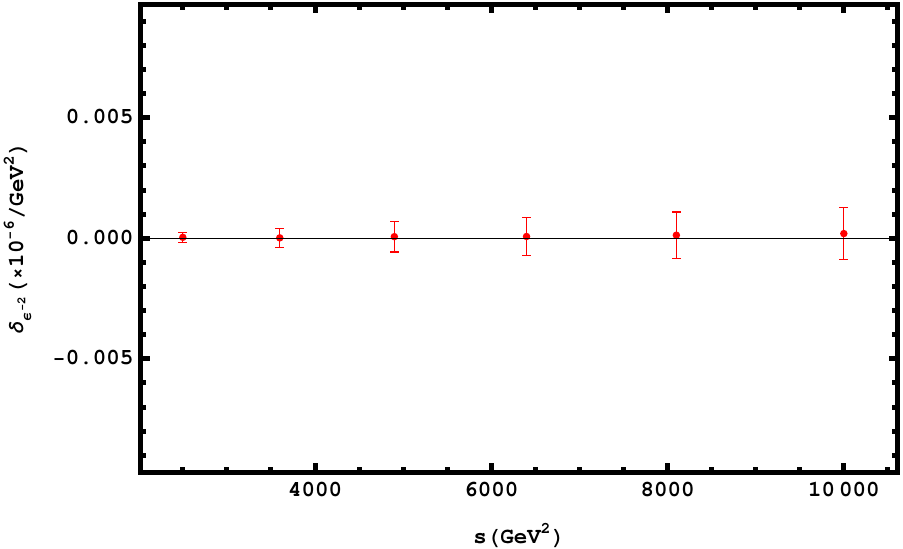}
\end{minipage}
\begin{minipage}[b]{3.2in}
  \includegraphics[width=3.2in,angle=0]{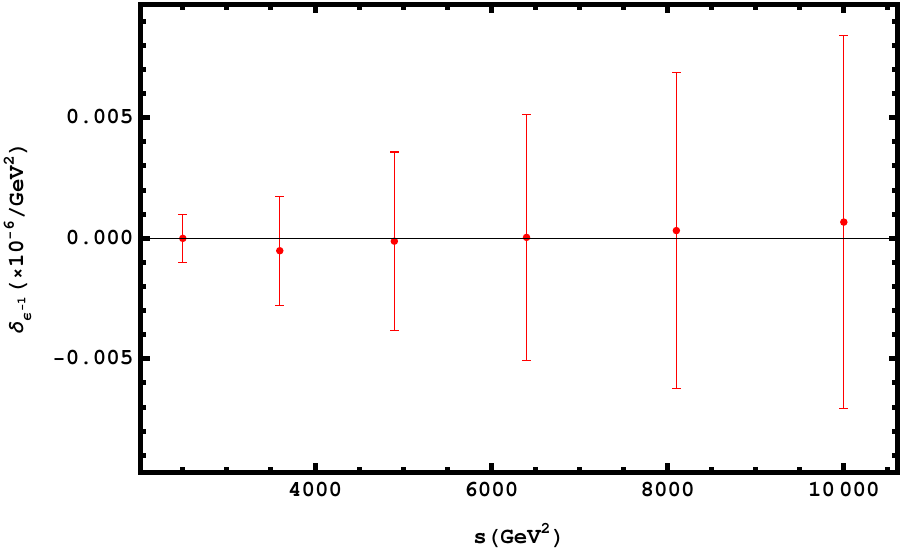}
\end{minipage}
\vspace{0.0cm}
\caption{Cancellation of $\epsilon^{-2}$ (left panel) and
$\epsilon^{-1}$ (right panel)
IR poles in Eqs.~(\ref{IRpole1})  and (\ref{IRpole2}) with numerical errors, as a function of machine center of mass energy squared {$s$}. This serves a non-trivial check on the consistency of our results.}
\label{e-poles}
\end{figure}

Note that all these IR pole terms are proportional to  $\delta(\tau_1)$. For the NLO virtual correction to the leading order process, this is easily understood since the final state consists of only a single colored particle ($q$) so that $\tau_1=0$. For the real emission contributions, the IR limit corresponds to either one or a pair of  particles being unresolved so that once again $\tau_1=0$. 

The condition in Eqs.~(\ref{IRpole1}) and (\ref{IRpole2}) must be satisfied for each value of the  final state jet transverse momentum ($P_{J_T}$) and rapidity ($y$).  In Fig.~\ref{e-poles}, we numerically demonstrate this cancellation of IR poles. In particular,  we plot sum of the coefficients of the $1/\epsilon^2$ (left panel) and $1/\epsilon$ (right panel) pole terms on the LHS of Eqs.~(\ref{IRpole1}) and (\ref{IRpole2}), after integrating over the Bjorken-$x$ variable, all allowed values of rapidity, and the jet transverse momentum $P_{J_T}$ with a lower cut of 20 GeV and  for a range of values of the center of mass energy $Q_e=\sqrt{s}$. The procedure of the extraction the IR poles for the real emission contributions in Eqs.(\ref{IRpole1}) and (\ref{IRpole2}) and their numerical computation is described in section \ref{nlocalc} and in the  appendices \ref{secNLO} and \ref{psdetail}.

\begin{figure}
\includegraphics[scale=1]{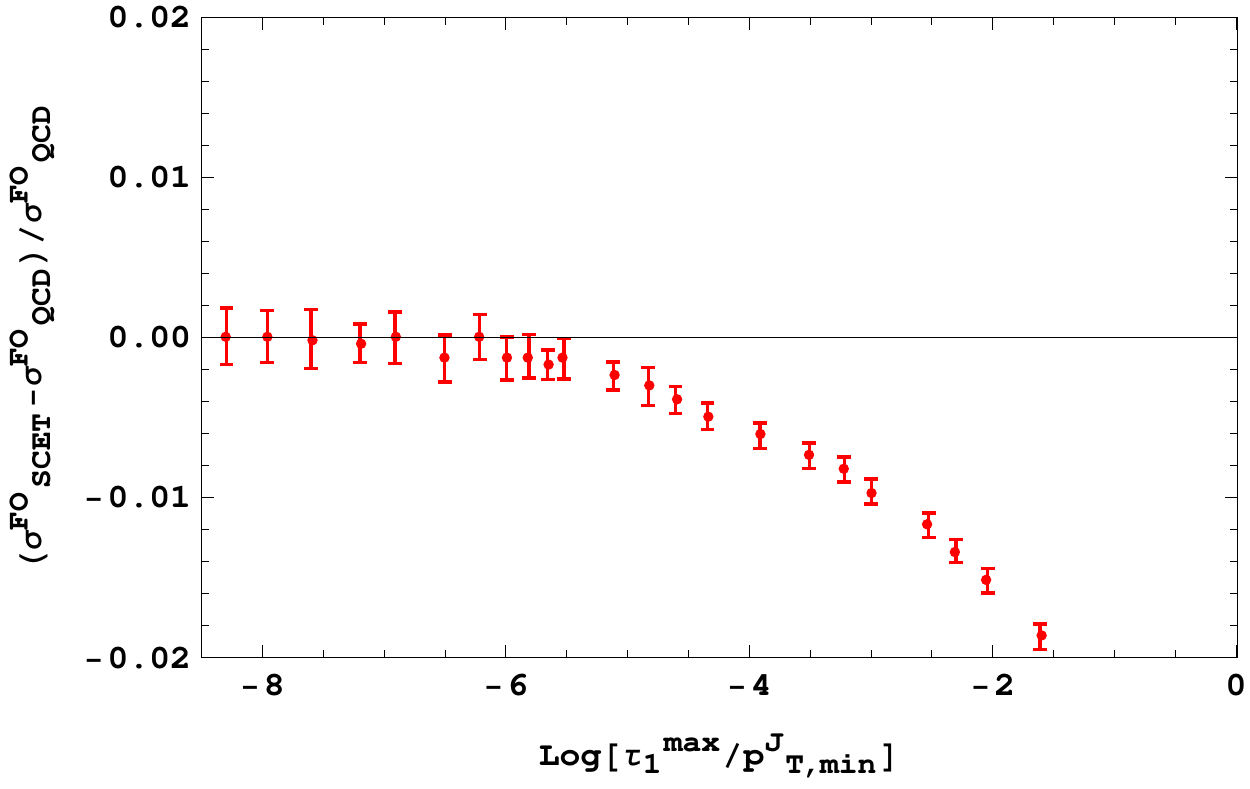}
\caption{In this plot, we compare the difference between the NLO QCD cross section
$\sigma_{\rm QCD}^{\rm FO}$ and the expanded SCET ${\cal O}(\alpha_s)$ 
prediction $\sigma_{\rm SCET}^{\rm FO}$, weighted to 
$\sigma_{\rm QCD}^{\rm FO}$.  The details on both cross sections are explained in
the text. In the resummation region where $\tau_1 \ll p_{J_T}$, the difference between 
these two predictions scales as $\tau_1/p_{J_T} \log^2(\tau_1/p_{J_T})$. Therefore as $\tau_1 \to 0$, the difference tends to $0$, as one can see from this plot, which implies that
 SCET correctly reproduces the singular  terms.}
\label{SCET-QCD}
\end{figure}

We also check that the fixed-order calculation in QCD reproduces the singular terms (in the $\tau_1\to 0$ limit) in the SCET result for the resummation region. In particular, in Eq.(\ref{match}), $d\sigma^{FO}$ must agree with $d\sigma_{\text{resum}}^{FO}$, the SCET resummed result expanded to fixed-order, in the $\tau_1 \to 0$ limit where the singular terms dominate. This is exactly the behavior reproduced in Fig.~\ref{SCET-QCD},  where we plot  the fractional difference between the QCD and expanded SCET results
\bea
\frac{\sigma_{\text{resum}}^{FO}(\tau_1^{\text{max}},P_{J_T}^{\text{min}})-\sigma^{FO}(\tau_1^{\text{max}},P_{J_T}^{\text{min}})}{\sigma^{FO}(\tau_1^{\text{max}},P_{J_T}^{\text{min}})},
\eea
as a function of $\text{Log}\>[\tau_1^{\text{max}}/P_{J_T}^{\text{min}}]$. The cross-sections $\sigma_{\text{resum}}^{FO}(\tau_1^{\text{max}},P_{J_T}^{\text{min}})$ and $\sigma^{FO}(\tau_1^{\text{max}},P_{J_T}^{\text{min}})$ are obtained by integrating $d\sigma_{\text{resum}}^{FO}$ and $d\sigma^{FO}$ over the $\tau_1$-range $[0,\tau_1^{\text{max}}]$, integrating over the final state jet momentum from $P_{J_T}^{\text{min}}$ to its maximum kinematically allowed value, and integrating over all kinematically allowed values of the jet rapidity $y$ for $Q_e=90$ GeV. In generating the plot of Fig.~\ref{SCET-QCD}, we set $P_{J_T}^{\text{min}}=20$ GeV and varied $\tau_1^{\text{max}}$. As expected, for $\tau_1^{\text{max}}/P_{J_T}^{\text{min}} \sim 1$ the expanded SCET and the fixed-order QCD computation results differ due to the non-singular (in the $\tau_1\to 0$ limit) terms but converge when $\tau_1^{\text{max}}/P_{J_T}^{\text{min}}\ll 1$, where the singular terms dominate. 

The full $\tau_1$ spectrum, has distinct regions 
\bea
\tau_1 &\sim & \Lambda_{QCD}, \nn \\
\Lambda_{QCD} \ll &\tau_ 1& \ll  P_{J_T}, \nn \\
\tau_ 1 &\sim & P_{J_T}, 
\eea
corresponding to the resummation region with non-perturbative soft radiation (soft function in Eq.(\ref{schem-1}) is non-perturbative), the resummation region with perturbative soft radiation, and the fixed-order region respectively. The two resummation regions are described by the factorization formula \cite{Kang:2012zr,Kang:2013wca} schematically described in Eqs.~(\ref{schem-1}) and (\ref{schem-2}). The factorization result depends on the scales $\mu_H,\mu_B,\mu_J,\mu_S$ with typical sizes given in Eq.~(\ref{scales}). These scales correspond to the natural scales that minimize large logarithms in the hard, beam, jet, and soft functions respectively. Using the renormalization group equations in the SCET, each of these functions is evolved to the common scale $\mu$ thereby summing large logarithms arising from ratios of the various disparate scales, corresponding to large logarithms of $\tau_1/P_{J_T}$. On the other hand, in the fixed-order region $\tau_1 \sim P_{J_T}$, the cross-section $d\sigma^{FO}$ is computed at a single common scale $\mu_{FO}$.
\begin{figure}
\includegraphics[scale=1.2]{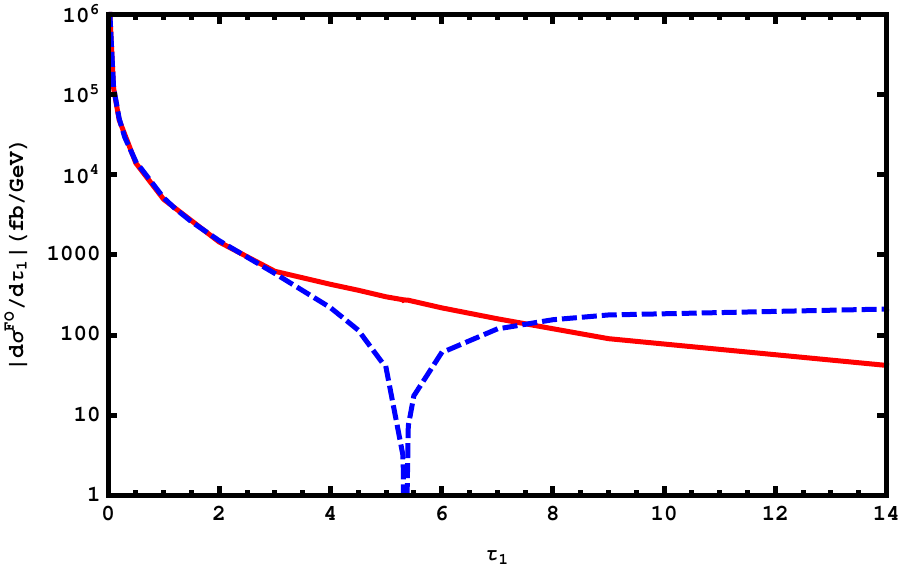}
\caption{We plot $|\mathrm{d}\sigma/\mathrm{d}\tau_1|$ for both full NLO QCD (red solid line) and the expanded
SCET singular (blue dashed line) predictions as a function of $\tau_1$. We see that when $\tau_1 \sim 5 {\rm GeV}$,
the singular contribution goes negative and the cross section is dominated by the nonsingular pieces which can not be predicted by SCET. This implies that the resummation should
start to be turned off around this  point and switched to the fixed-order QCD prediction
smoothly. This figure
justifies the parameters we choose for the profile-scales for matching as explained in the text.}
\label{tau1-SCET-QCD}
\end{figure}

In Fig.~\ref{tau1-SCET-QCD}, we show $|\mathrm{d}\sigma/\mathrm{d}\tau_1|$ for the expanded (resummation turned off) SCET (dashed blue) and the fixed-order QCD (solid red) results. The quantity $|\mathrm{d}\sigma/\mathrm{d}\tau_1|$ in Fig.~\ref{tau1-SCET-QCD} was obtained by integrating $d\sigma$, as defined in Eq.(\ref{obs}), over a the final state jet transverse momentum and rapidity in the range  $[P_{J_T}^{\text{low}},P_{J_T}^{\text{high}}]=[20\>{\rm GeV}, 30\>{\rm GeV}]$ and $|y| < 2.5$ respectively. The  scales were chosen to be $\mu^{FO}=\mu=\mu_H=\mu_B=\mu_J=\mu_S=2P_{J_T}^{\text{low}} $. As expected in the limit $\tau_1 \to 0$, the expanded SCET and the fixed-order QCD results converge, since this region is dominated by the singular terms in the cross-section, correctly reproduced by SCET. For larger values of $\tau_1$, the non-singular terms not reproduced by SCET, become important and the expanded SCET and fixed-order QCD results differ. From Fig.~\ref{tau1-SCET-QCD}, we see that for $\tau_1 \sim 5$ GeV, the expanded SCET cross-section goes negative implying that the non-singular terms in the fixed-order QCD result cannot be ignored. This suggests that in matching the SCET resummed and fixed-order QCD results (see Eq.(\ref{match})), the SCET resummation should start to be turned off around $\tau_1\sim 5$ GeV. In this manner, one can determine the appropriate regions in which resummation of the SCET result can be turned off and matched onto the fixed-order QCD result. 

\subsection{Scale Variation and Profile Functions}

As described in Refs.~\cite{Berger:2010xi,Stewart:2011cf}, care must be taken to properly estimate the perturbative uncertainties in the calculation of the matched cross-section in Eq.(\ref{match}). The region $\tau_1\ll P_{J_T}$, corresponds to configurations with a single narrow jet and only soft radiation at wide angles, as seen in the left panel of Fig.~\ref{fig:process}. If one integrates the cross-section over $\tau_1$ in the range $[0,\tau_1^{\text{cut}}]$, then $\tau_1^{cut}$ acts as a jet veto parameter. A small value of $\tau_1^{cut}$ corresponds to a strong veto on additional jets. It amounts to dividing the total cross-section into an exclusive 1-jet bin and a 2-jet inclusive cross-section. As one increases $\tau_1^{cut}$,  the jet veto is relaxed and additional hard radiation or jets are allowed at wide angles, and one approaches the total inclusive cross-section.   Thus, the analysis of the perturbative uncertainty in the presence of $\tau_1^{cut}$, is similar to that in exclusive single jet production where  correlations between the perturbative uncertainty in the 1-jet bin cross-section and the total cross-section must be taken into account. In particular, a direct scale variation of the fixed-order calculation of the 1-jet bin cross-section, obtained through imposing a small $\tau_1^{\rm cut}$, underestimates the  uncertainty due to an absence of the correlation information. 

\begin{figure}
\subfigure [] {\includegraphics[scale=0.62]{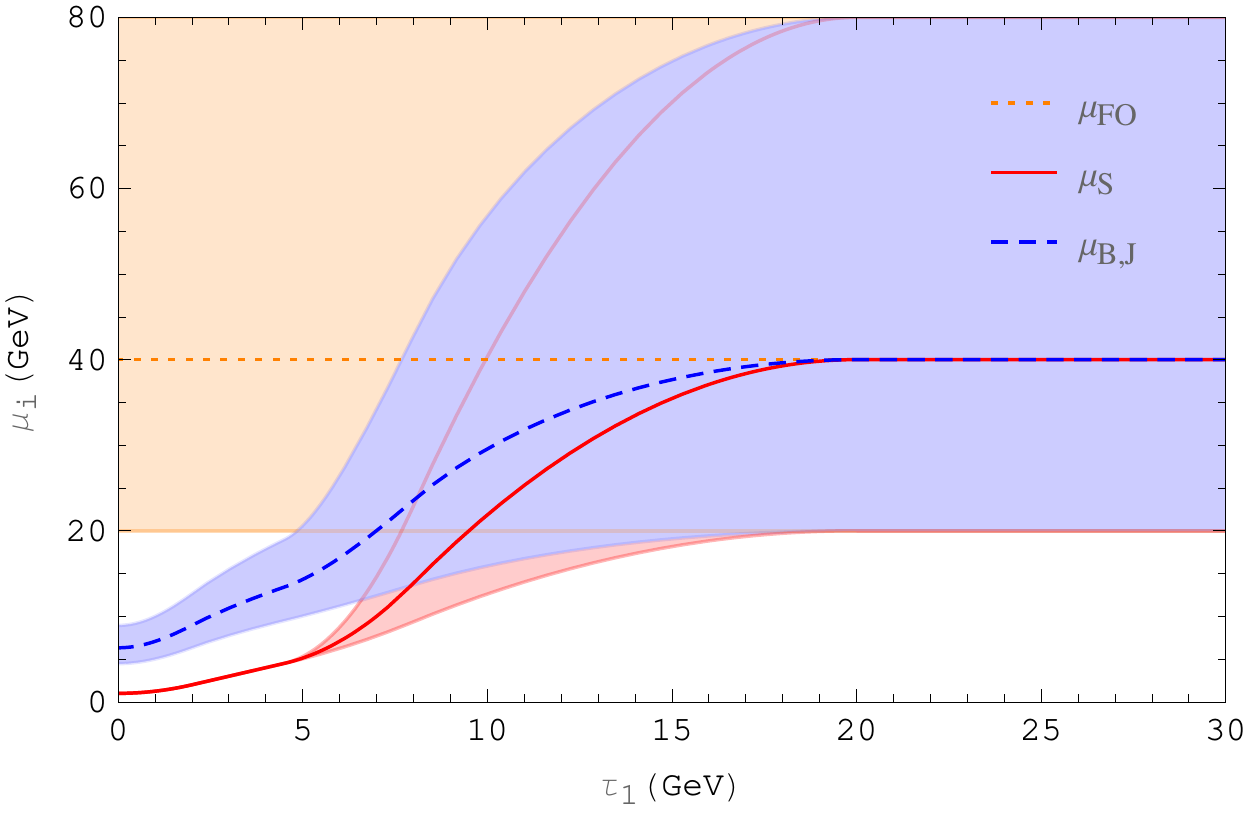}}
\subfigure [] {\includegraphics[scale=0.62]{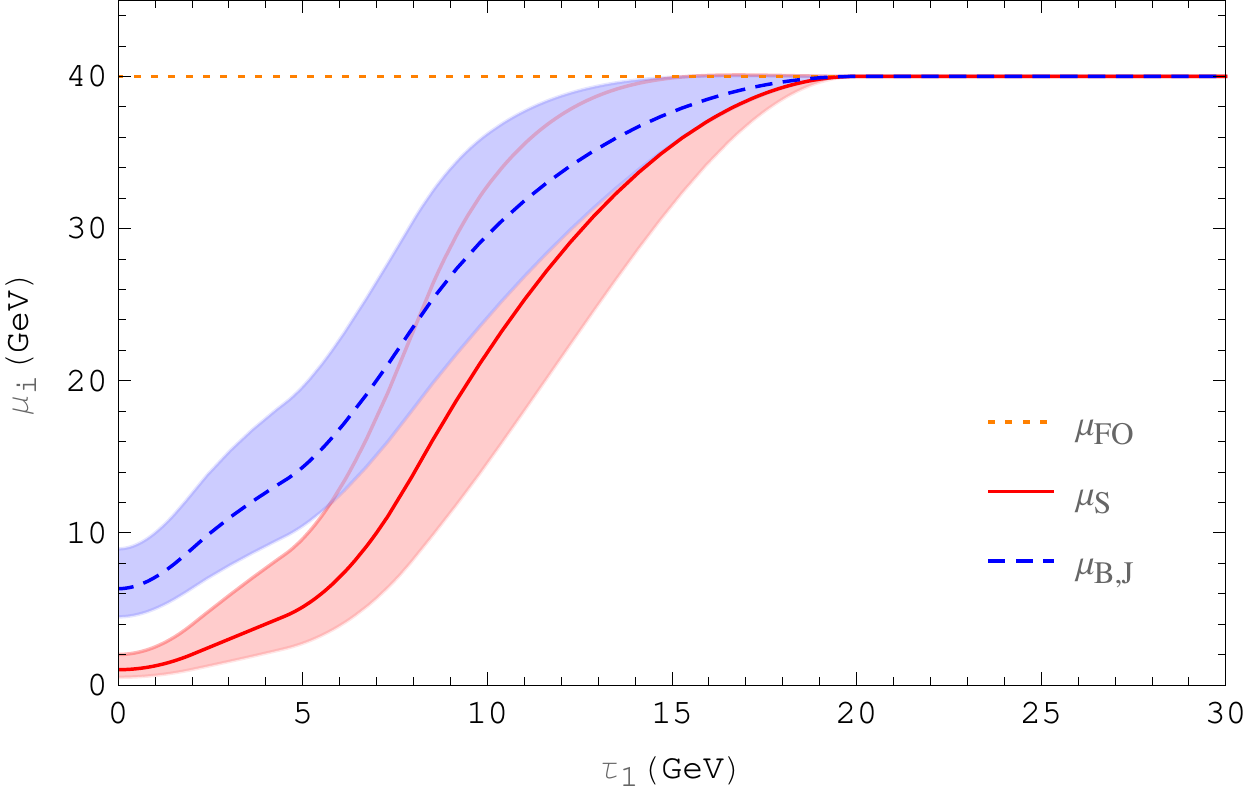}}
\caption{Profile functions for the SCET scales $\mu_{S,B,J}$ are chosen so that they converge to $\mu_{FO}$ in the large $\tau_1$ region where perturbarive QCD is appropriate. The left panel shows the collective scale variation defined through Eqs.~(\ref{profile1}) and (\ref{muFO}). The right panel shows the SCET resummation scale variation defined through Eqs.~(\ref{profile1}) and (\ref{muSCET}), corresponding to jet binning uncertainties. The total scale variation uncertainty is given by adding these two types of scale variation uncertainties in quadrature as in Eq.(\ref{band}).}
\label{varyscale}
\end{figure}

As discussed in Refs. \cite{Berger:2010xi,Stewart:2011cf}, the perturbative uncertainties with jet binning correlation information can be reliably estimated using the framework for resummation and fixed-order matching in Eq.(\ref{match}), along with profile functions for the SCET scales $\mu_H,\mu_B,\mu_J,\mu_S,$ and $\mu$ in $d\sigma_{\text{resum}}$. In the fixed-order calculation $d\sigma^{FO}$, both the singular logarithmic terms and the non-singular terms are evaluated at the common scale $\mu_{FO}$. Thus, in order to match the resummation regions and the fixed-order regions, resummation must be turned off in the region $\tau_1\sim P_{J_T}$ so that the scales $\mu_H,\mu_B,\mu_J,\mu_S,$ and $\mu$  smoothly converge to $\mu_{FO}$, as accomplished by using profile functions (see Fig.~\ref{varyscale}). This  ensures that important cancellations that occur between the singular and non-singular terms  in the fixed-order region $\tau_1 \sim P_{J_T}$ region, are correctly reproduced. In addition, the perturbative uncertainty in the fixed-order region is given by a scale variation of a single scale $\mu_{FO}$ in the fixed-order cross-section. Since the profile functions ensure that various the SCET scales smoothly converge to $\mu_{FO}$ in the fixed-order region, the correct perturbative uncertainty is reproduced in the fixed-order region.  In the resummation region, corresponding to a strong jet veto, the correlation uncertainties between the 1-jet bin and the total cross-section are reproduced~\cite{Berger:2010xi,Stewart:2011cf} by  scale variations of $\mu_B,\mu_J,$ and $\mu_S$. As one increases $\tau_1^{cut}$ to large values, the 1-jet bin cross-section dominates the 2-jet inclusive cross-section so that the jet bin correlation is no longer important. This  is reproduced by the fact that the profile functions~\cite{Ligeti:2008ac,Abbate:2010xh} smoothly turn off resummation in the large $\tau_1^{cut}$ region and the scale variation amounts to the variation of the single scale $\mu_{FO}$. In the context of DIS, such profile functions were first used in Ref.~\cite{Kang:2013nha} for their NNLL results in the resummation region.

We choose profile functions, following Ref.~\cite{Stewart:2013faa}, for the scales $\mu_{FO}, \mu, \mu_H, \mu_B,\mu_J$, and $\mu_S$ as
\bea
\label{profile1}
&&\mu_{FO}=\mu_H=\mu = 2p_{J_T} \,, \nn \\
&&\mu_S = \mu \, f_{\rm profile}(\tau_1/\mu) \,, \nn \\
&&\mu_B = \mu_J = \sqrt{\mu\> \mu_S}\,,
\eea
where $ f_{\rm profile}(\tau_1/\mu)$ denotes the profile function which ensures that the scales $\mu_B,\mu_J$, and $\mu_S$ smoothly approach $\mu_{FO}$ in the fixed-order region. The profile function  $ f_{\rm profile}(\tau_1/\mu)$  is built by choosing five distinct functions for five corresponding  regions of the $\tau_1$ spectrum, such that it's continuous and has  continuous derivatives across all five regions. More specifically, we choose \cite{Stewart:2013faa}
\begin{align} \label{profile2}
f_{\rm profile} (x) =
\begin{cases}
x_0 \bigl[ 1 + (x / x_0)^2/4 \bigr] & x \le 2x_0 \,, \qquad \qquad\> \text{Region I}\, \\
x & 2x_0 \le x \le x_1 \,,  \qquad \text{Region II}\,\\
x + \dfrac{(2 - x_2 - x_3) (x - x_1)^2}{2(x_2 - x_1) (x_3 - x_1)} & x_1 \le x \le x_2 \,, \qquad \>\>\text{Region III} \vspace{1ex} \\
1 - \dfrac{(2 - x_1 - x_2) (x - x_3)^2}{2(x_3 - x_1) (x_3 - x_2)} & x_2 \le x \le x_3 \,, \qquad\>\> \text{Region IV} \\
1 & x_3 \le x \, \qquad \qquad\>\>\>\>\>\>\text{Region V}.
\end{cases}
\end{align}
where the parameters $x_{0,1,2,3}$ are chosen to have the values 
\bea
\label{profile3}
x_0 = 1\>{\rm GeV}/\mu, \qquad
x_1 = 4.5\>{\rm GeV}/\mu,   \qquad
x_2 = 8.0\>{\rm GeV}/\mu,  \qquad
x_3 = p_{J_T}^{\rm low}/\mu.
\eea
In region I, resummation of large Sudakov logarithms is required and the soft function is sensitive to non-perturbative effects which must be modeled. We implement the same model for the soft function used in Refs.~\cite{Kang:2012zr,Kang:2013wca}, based on the method used in Refs.~\cite{Ligeti:2008ac,Hoang:2007vb}. Region II corresponds to the resummation region with a perturbative soft function. Region V corresponds to the fixed-order region where all the resummation scales converge to $\mu_{FO}$. Regions III and IV are the intermediate transition regions between the resummation and fixed-order regions.  Fig.~\ref{varyscale} shows the profiles for $\mu_H,\mu_B,\mu_J$, and $\mu_S$ and their convergence to $\mu^{FO}$ in the fixed-order region. It also shows the scale variation employed in our analysis, as discussed later in this section.

\begin{figure}
\includegraphics[scale=1]{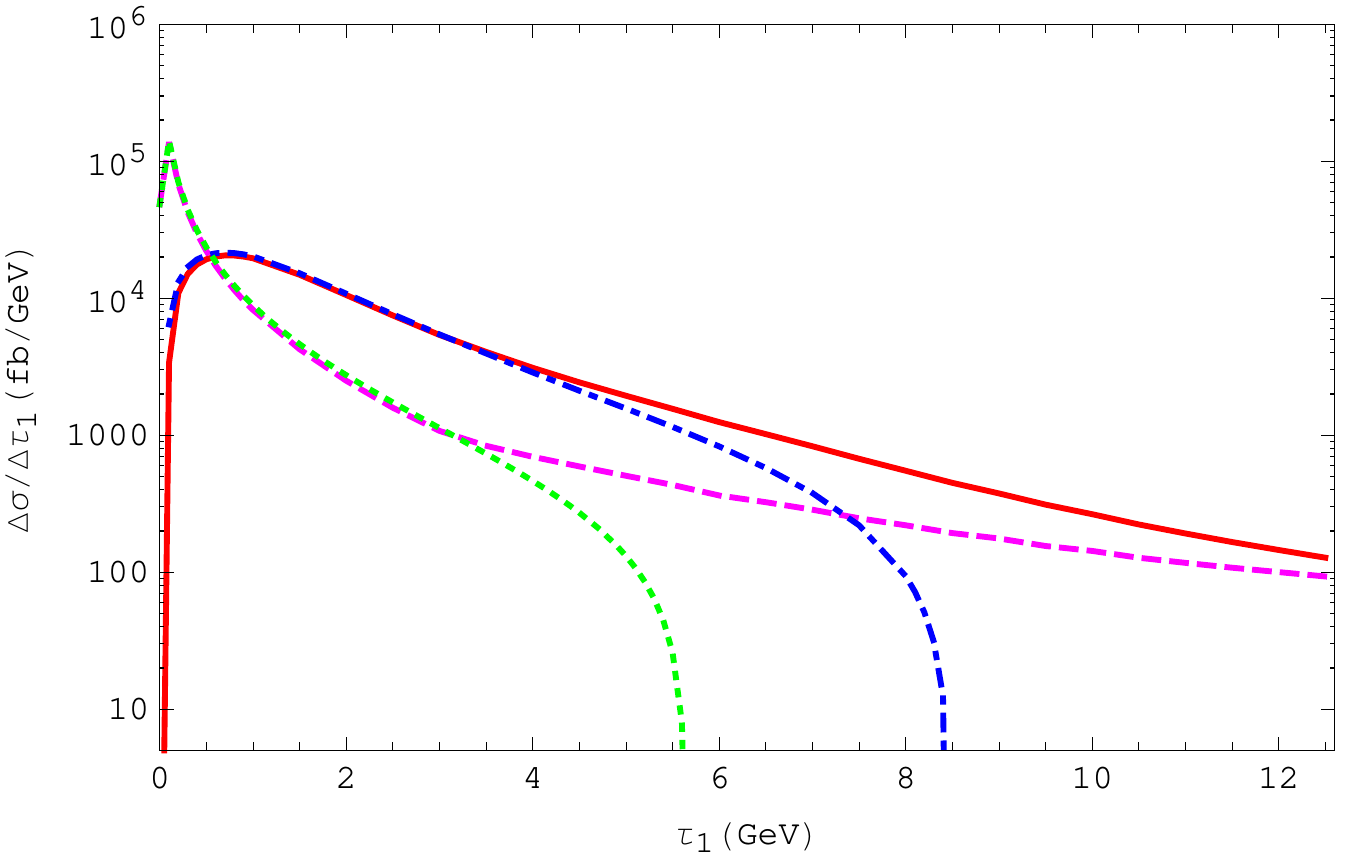}
\caption{Here we show the $\tau_1$-distribution for the observable in Eq.~(\ref{obs}) with a proton target on EIC predicted by NNLL $+$ NLO QCD (red solid), NLO QCD only (dashed magenta), expanded SCET (green dotted) and NNLL only (dot-dashed blue).}
\label{eictau1}
\end{figure}

In Fig.~\ref{eictau1}, we give results for the $\tau_1$ distribution for the observable in Eq.~(\ref{obs}) for a proton target at a machine center of mass energy of $90 \>{\rm GeV}$. We have integrated Eq.~(\ref{obs}) over $P_{J_T}$ in the range $[P_{J_T}^{\text{low}},P_{J_T}^{\text{high}}]=[20\>{\rm GeV}, 30\>{\rm GeV}]$ and 
restricted the jet rapidity to $|y| < 2.5 $. The distribution in $\tau_1$ is obtained by integrating
$\tau_1$ over bins of size $0.1\> {\rm GeV}$ and dividing the result  by the same bin size. For the fixed-order ($d\sigma^{FO}$) and expanded SCET ($d\sigma^{FO}_{\text{resum}}$)  results, we choose $\mu_{FO}=2P_{J_T}^{\text{low}}$.  For the NNLL resummed contribution ($d\sigma_{\text{resum}}$), the scales are set as in Eqs. (\ref{profile1}), (\ref{profile2}), and (\ref{profile3}). We use CTEQ6m PDFs and follow the conventions in table 1 of Ref.~\cite{Berger:2010xi} for the order of $\alpha_s$ running, the order of PDFs and matrix elements, and the counting of logarithms. In particular,  we use $2$-loop $\alpha_s$ running and NLO PDFs for the fixed-order calculations and  $3$-loop $\alpha_s$ running and NLO PDFs for the NNLL resummation contribution. 
We see that, as expected, the NNLL+NLO matched  curve (solid red) converges with the NNLL resummation curve (dashed-dot blue) for small $\tau_1$ and converges towards the fixed-order NLO curve (dashed magenta) for large $\tau_1$. As before, we also see that the expanded SCET (dotted green) and the fixed-order NLO (dashed magenta) curves converge for small values of $\tau_1$. 

Note that, as mentioned earlier, for $\tau_1\sim \Lambda_{QCD}$ the soft function is non-perturbative and must be modeled. However, these non-perturbative effects are not included in the curves shown in Fig.~\ref{eictau1}. Instead the curves are generated using the purely perturbative expression for the soft function. The results with a non-perturbative model for the soft function will be presented later in this section. The focus of Fig.~\ref{eictau1} is  to demonstrate that the NNLL+NLO curve converges to the NNLL curve in the resummation region and approaches the NLO curve in the fixed-order region, as expected.

The curves in Fig.~\ref{eictau1} were generated for central values of the various renormalization scales. We now turn to scale variation to estimate the perturbative uncertainties. The perturbative uncertainty from the NLO contribution $d\sigma^{FO}$ (see Eq.~(\ref{match})) is estimated in the standard way by varying the single scale $\mu_{FO}$, where we have set the renormalization ($\mu_r$) and factorization ($\mu_F$) scales equal to each other so that $\mu_r=\mu_F=\mu_{FO}$. The same variation is applied to the expanded SCET contribution $d\sigma_{\text{resum}}^{FO}$, defined in Eq.~(\ref{resumoff}). In our numerical analysis we vary $\mu_{FO}$ by a factor of 2 about its central value 
\bea
\label{muFO}
\mu_{FO}^{\text{up}} = 2 \>\mu_{FO}^{\text{central}}, \qquad \mu_{FO}^{\text{down}} = \mu_{FO}^{\text{central}}/2, \qquad \mu_{FO}^{\text{central}} = 2 P_{J_T}^{\text{low}},
\eea
so that $\mu_{FO}^{\text{up}}$ and $\mu_{FO}^{\text{down}}$ define the scale variation envelope.

The second set of scale variations, related to correlations between the total and the 1-jet bin cross-sections described earlier, is given by independent variations of the scales $\mu_J,\mu_B,\mu_S$ in the resummed contribution $d\sigma_{\rm resum}$. As discussed earlier, these scales are given by the profile functions according to Eqs.~(\ref{profile1}) and (\ref{profile2}) in order to ensure that resummation is smoothly turned off in the fixed-order region.  In particular, we vary the scales as 
\bea
\label{muSCET}
&&\mu_{B/J}^{\rm up} = \mu_{B/J}^{\rm central} \times \sqrt{f_{\rm var}}, \qquad
\mu_{B/J}^{\rm down} = \mu_{B/J}^{\rm central} / \sqrt{f_{\rm var}}  \,, \nn \\
&&\mu_{S}^{\rm up} = \mu_{S}^{\rm central} \times {f_{\rm var}}, \qquad \mu_{S}^{\rm down} = \mu_{S}^{\rm central} / {f_{\rm var}} \,,
\eea
where the variation profile function is given by
\begin{align} \label{eq:fvar}
f_{\rm var} (x) =
\begin{cases}
2(1-x^2/x_3^2) & x \le 2x_0 \,, \\
1+2(1-x/x_3)^2 & x_3/2 \le x \le x_3 \,, \\
1 &  x \le x_3 \,,
\end{cases}
\end{align}
and as before the ``up" and ``down"  scales define the envelope of scale variations.
This prescription follows the analysis of Ref.~\cite{Stewart:2013faa}, but is adapted to our framework where the see-saw relation between the hard, jet, and soft scales must be maintained 
\bea
\mu_J^2 \sim \mu_B^2 \sim \mu_H \mu_S.
\eea
We note that, as required, the variation profile function $f_{\rm var}$ smoothly approaches unity in the fixed-order region $x>x_3$, corresponding to the fixed-order region where resummation and the associated scale  variations must turn off. The total scale variation uncertainty is schematically given by the sum of two contributions added in quadrature
\bea
\label{band}
\Delta^2_{{\rm NNLL} + {\rm NLO} } =\,
\Delta^2_{\rm FO} + \Delta^2_{\rm resum} \,,
\eea
since they are uncorrelated.
The uncertainty  $\Delta_{\rm FO}$ is associated with the scale variation of $\mu_{FO}$ and corresponds to an overall variation of all scales by the same factor, as seen in Eq.(\ref{profile1}). It corresponds to the usual scale variation in fixed-order calculations which becomes explicit in the fixed-order region where all scales approach the common value $\mu_{FO}$, according to their profile functions. The uncertainty $\Delta_{\rm resum}$ corresponds to the independent variation of $\mu_{J,B,S}$ with $\mu_{FO}$ fixed at its central value. As discussed earlier, these resummation uncertainties are associated with uncertainties arising from jet-bining. Another source of scale variation can be obtained by varying the parameters $x_{0,1,2,3}$ in Eq.(\ref{profile3}), which determine the transition regions of the profile functions. Typically, the uncertainties associated such variations are much smaller than the other scale variations discussed~\cite{Stewart:2013faa}.  For more details of this type uncertainty analysis for exclusive jet processes, we refer the reader to Refs.~\cite{Berger:2010xi,Stewart:2011cf,Stewart:2013faa}

\subsection{Non-perturbative Soft Radiation}

Before presenting results with scale variation uncertainties, we remind the reader that in the region $\tau_1\sim \Lambda_{QCD}$, the soft function in $d\sigma_{\rm resum}$  (see Eq.~(\ref{schem-1})) becomes non-perturbative. Thus, a complete description of the $\tau_1$ spectrum requires a non-perturbative model for the soft function. The model should reduce to the perturbative result as one increases $\tau_1$ into the perturbative region, for a smooth matching between the perturbative and non-perturbative regions. We now briefly describe the elements of the soft function model used in the numerical analysis. For more details and complete field-theoretic definitions, we refer the reader to Ref.~\cite{Kang:2013wca}. 

The soft function appearing in the factorization formula for $d\sigma_{\rm resum}$ can be written as
\bea
{\cal S}(\tau_1,\mu_S) &=& \int dk_a \int dk_J\> \delta(\tau_1-k_a-k_J){\cal S} (k_a,k_J,\mu_S),
\eea
where ${\cal S} (k_a,k_J,\mu_S)$ appearing on the RHS is the generalized hemisphere soft function \cite{Jouttenus:2011wh}. The arguments $k_a,k_J$ correspond to the contribution to $\tau_1$ of the soft radiation grouped with the beam and jet directions respectively.
One can write a non-perturbative model for the soft function by writing the generalized hemisphere soft function as a convolution~\cite{Ligeti:2008ac,Hoang:2007vb} of the partonic soft function ${\cal S}_{\text{part.}}$, calculated in perturbation theory,  and a model function $S_{\text{mod.}}$ 
\bea
{\cal S} (k_a,k_J,\mu_S) &=& \int dk_a' \int dk_J' \> {\cal S}_{\text{part.}} (k_a - k_a,k_J-k_J', \mu_S) S_{\text{mod.}} (k_a',k_J'),
\eea
where the model function $S_{\rm mod.}$ satisfies the normalization condition
\bea
 \int dk_a' \int dk_J' \>S_{\text{mod.}} (k_a',k_J') &=& 1.
\eea
Such a model correctly reproduces the scale dependence of the soft function through the perturbatively calculable ${\cal S}_{\text{part.}}$. The model function $S_{\text{mod.}}$ is chosen to have a peak around $k_{a,J}'\sim \Lambda_{QCD}$.  This allows for an operator product expansion in the region $\tau_1 \gg \Lambda_{QCD}$, such that the leading term reduces to the perturbatively calculable partonic soft function, as required.
As described  in Refs.~\cite{Kang:2012zr,Kang:2013wca}, the factorization formula in Eq.(\ref{schem-1}) depends on $S_{\rm mod.}$ through the combination
\bea
F_{\rm mod.} (u) &=& \int_{-u}^u \frac{d\zeta}{2} \> {\cal S}_{\text{mod.}} (\frac{u+\zeta}{2},\frac{u-\zeta}{2}), 
\eea
where $u$ and $\zeta$ are related to the original variables as $u=k_a'+k_J', \zeta= k_a'-k_J'$ and $F_{\rm mod.}$ satisfies the  normalization condition
\bea
\label{normF}
\qquad \int_0^\infty du \> F_{\rm mod.}(u) =1.
\eea
For our numerical analysis, we use a model of the form
\bea
\label{Fmod}
F_{\rm mod.} (u) = \frac{N(a,b,\Lambda)}{\Lambda} \left ( \frac{u}{\Lambda} \right )^{a-1} \> \text{Exp} \Big[ -\frac{(u-b)^2}{\Lambda^2}\Big ] ,
\eea
where $a,b,\Lambda$ correspond to parameters of the model and $N(a,b,\Lambda)$ is a normalization factor chosen to satisfy the condition in Eq.~(\ref{normF}). Finally, we note that the case of the purely perturbative soft function with no model corresponds to the choice $S_{\rm mod.}(k_a,k_J)=\delta(k_a)\delta(k_J)$ or equivalently $F_{\rm mod.}(u)=\delta(u)$. For further details, we refer the reader to Ref.~\cite{Kang:2012zr,Kang:2013wca}. 
\begin{figure}
\includegraphics[scale=1]{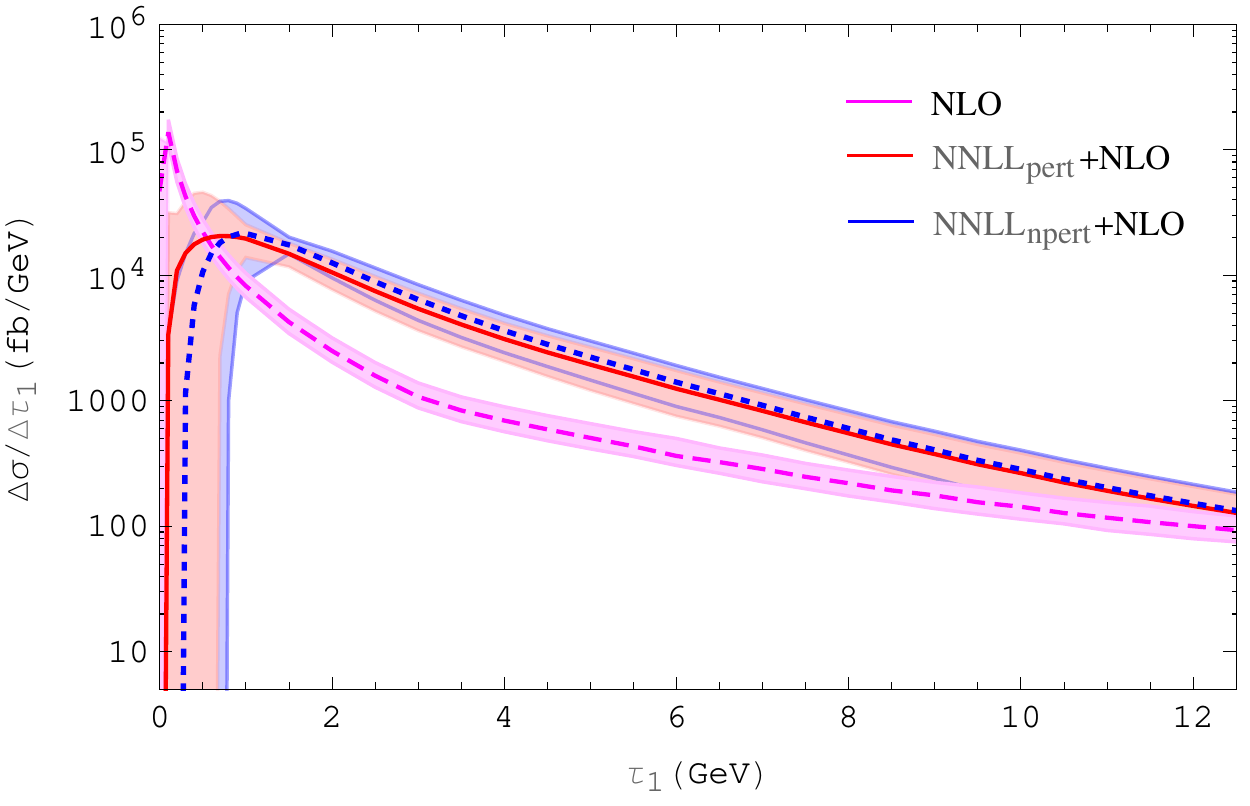}
\caption{Here we show the full spectrum of the $\tau_1$ distribution predicted by
NLO (dashed magenta), ${\rm NNLL}+{\rm NLO}$ (solid red) and  ${\rm NNLL}+{\rm NLO}$ 
convoluted with
a non-perturbative soft function model (dotted blue). The 
scale uncertainty bands are estimated using the scheme described in the text.
}
\label{un}
\end{figure}

In Fig.~\ref{un}, we show the numerical results for the complete NNLL+NLO $\tau_1$-spectrum. The various curves are generated for a proton target with $Q_e=90$ GeV, the final state jet transverse momentum is integrated over the range  $[P_{J_T}^{\text{low}},P_{J_T}^{\text{high}}]=[20\>{\rm GeV}, 30\>{\rm GeV}]$, the jet rapidity is integrated over the range $|y| \leq 2.5$, and we have integrated over $\tau_1$ bins of size 0.1 GeV and divided by this bin size. We used the anti-$k_T$ jet algorithm with jet radius  $R=1.0$ in order to determine the jet reference vector $q_J$ in the calculation of the fixed-order contribution $d\sigma^{FO}$ in Eq.(\ref{match}). The solid red curve corresponds to the NNLL+NLO $\tau_1$-spectrum for the central values of the scales as described in Eqs.~(\ref{profile1}) and (\ref{muFO}) and is identical to the solid red curve in Fig.~\ref{eictau1}. It corresponds to the case of a purely perturbative soft function, corresponding to the choice $F_{\rm mod.}(u)=\delta(u)$. Here we also show the scale variation uncertainty band (pink band), generated using Eq.~(\ref{band}). The dashed blue curve incorporates a non-perturbative model for the soft function and is generated for the same central scale choices as the solid red curve. For the non-perturbative soft function, the model parameters were chosen to be $a=1.8,b=-0.05\>{\rm GeV} , \Lambda=0.4 \>{\rm GeV}$ for $F_{\rm mod.}$ in Eq.~(\ref{Fmod}). The blue band gives the corresponding scale variation uncertainty as determined by Eq.~(\ref{band}). For comparison, we also show the curve (dashed magenta) for the NLO contribution $d\sigma^{FO}$ and its scale variation uncertainty band, resulting from the variation in Eq.~(\ref{muFO}). We see that the NNLL+NLO matched curves converge to the pure NLO curve for large $\tau_1$ as expected.

\begin{figure}
\includegraphics[scale=1]{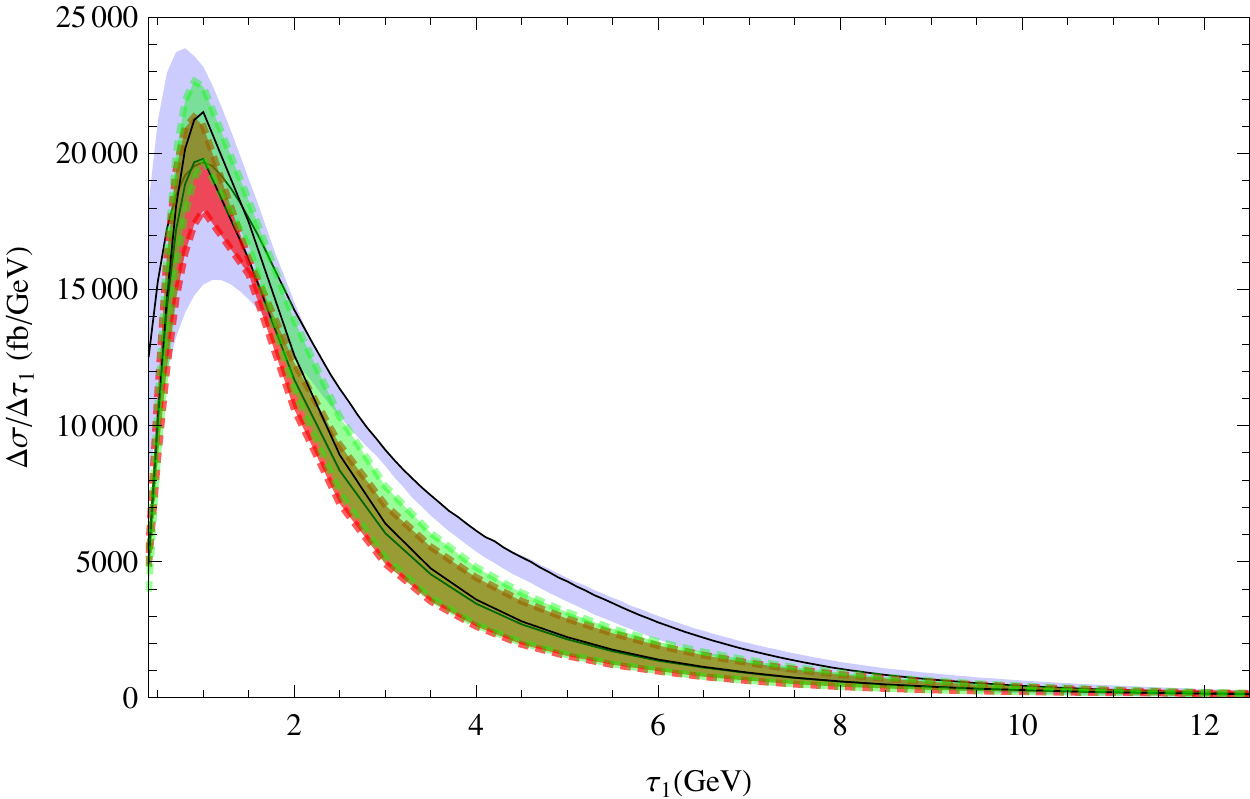}
\caption{Here we show a comparison of the NLL (widest blue band), NLL$'$+NLO (red band) and the NNLL+NLO (green band) $\tau_1$-spectrum.  The non-perturbative soft function model is the same as described in the text and used in Fig.~\ref{un}. We see an overall reduction in the theoretical uncertainty as we include higher order resummation effects.
}
\label{convergence}
\end{figure}

We note that non-perturbative effects from soft function distort the $\tau_1$-spectrum primarily in the region $\tau_1\sim \Lambda_{QCD}$, as seen by the difference in the solid red and blue dashed curves in Fig.~\ref{un}, corresponding to the curves with and without a non-perturbative soft function model respectively. For $\tau_1\gg \Lambda_{QCD}$, the two curves converge, indicating that the non-perturbative model smoothly turns off in the perturbative region as required. The parameters of the soft function model can be extracted through a fit to data. We note that even in the region $\tau_1\gg \Lambda_{QCD}$, the model soft function can affect the distribution through a power correction, determined by the first moment~\cite{Salam:2001bd,Lee:2006fn,Lee:2006nr,Abbate:2010xh,Mateu:2012nk,Kang:2013nha} of the model function. As discussed in Refs.~\cite{Kang:2012zr,Kang:2013wca}, this soft function is universal, being independent of the nuclear target. Thus, the soft function can be measured from data off a proton target and used as a known quantity for other nuclear targets. We point out that the phenomenological  model soft function used here is for illustrative purposes only. i.e. to demonstrate that the formalism allows for the implementation of a soft function model that can be smoothly connected with the spectrum in the perturbative region and  correctly reproduces the scale dependence of the soft function. Determining the true impact of the non-perturbative soft radiation requires comparison with data. 

In Fig.~\ref{convergence}, we also show a comparison of the NLL, NLL$'$+NLO and the NNLL+NLO matched $\tau_1$ spectra, using the same illustrative soft function model. NLL$'$ is defined as NLL resummation with the singular part of the NLO matrix element. For a detailed summary of these conventions, see Table ${\rm I}$ of Ref.~\cite{Berger:2010xi}. We see an overlap of the scale variation uncertainty bands and an overall reduction in theoretical uncertainty as we include higher order resummation effects.

\begin{figure}
\includegraphics[scale=1]{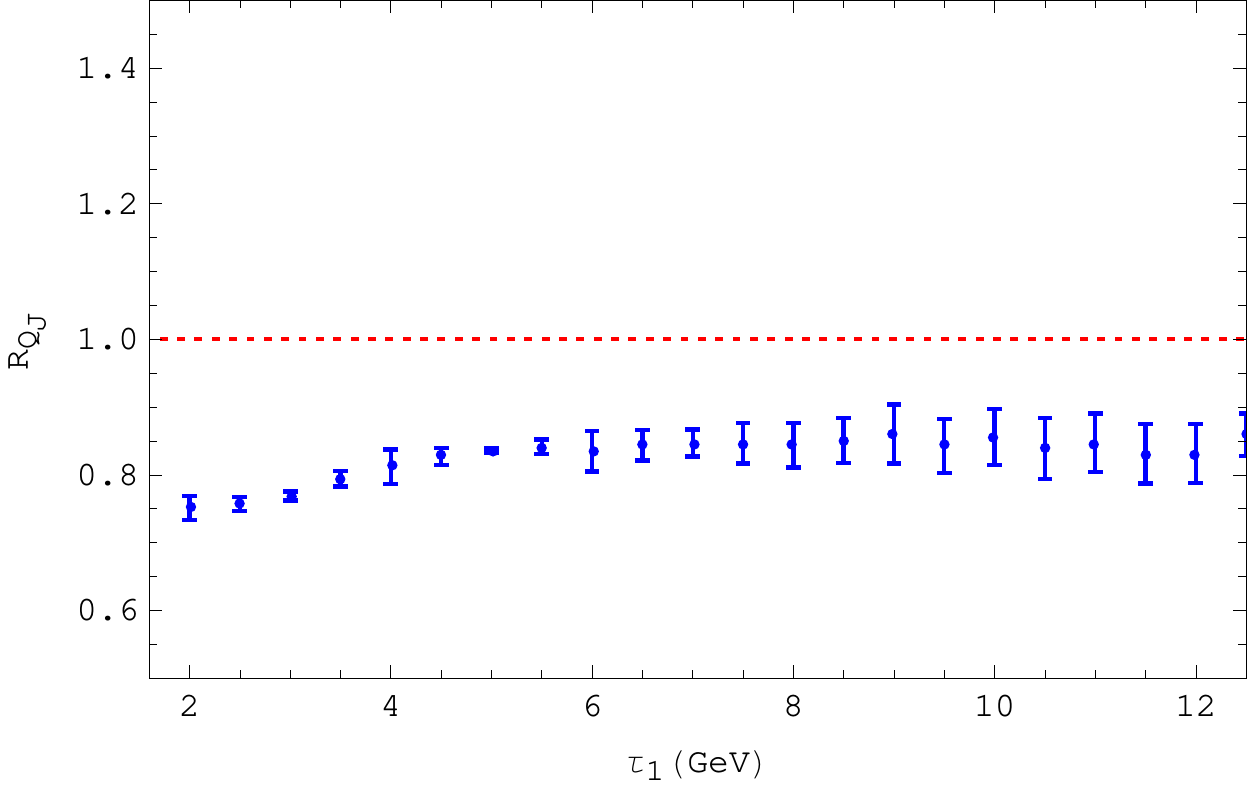}
\caption{Here we show the ratio of $\tau_1$-distributions for two different choices of $Q_J$ appearing in the definition of  $\tau_1$ in Eq.~(\ref{1-jettiness}) and the jet momentum $P_J$ defined in Eq.(\ref{pjet}). In particular, we plot the ratio of the distribution with $Q_J=2K_{J_T}\cosh y_K$ to that with $Q_J=2\rho(R,y_K)K_{J_T}\cosh y_K$ where $\rho(R,y_K)$ allows one to change the size and shape of the jet region. We make the choice~\cite{Jouttenus:2013hs} $\rho(R=1,y_K)=0.834-0.233y_K^2+0.077 y_K^4-0.008 y_K^6$.}
\label{QJshape}
\end{figure}

\subsection{Jet Shape Analysis}

The 1-jettiness framework also allows for a jet-shape analysis. One can study changes to the $\tau_1$-distribution for different definitions of the final state jet momentum, corresponding to the amount of radiation included in the jet. This can be a powerful tool for studying jet energy loss mechanisms. When such an analysis is applied for a variety of nuclear targets, it can serve as a probe of jet quenching or energy loss of fast partons moving through the nuclear medium. Here we only give numerical results for the proton target at the NNLL+NLO accuracy and leave the analysis of heavier nuclear targets for future work. For NNLL resummation results for a variety of nuclear targets in the region $\tau_1\ll P_{J_T}$, we refer the reader to the earlier work in Ref.~\cite{Kang:2013wca}.

The jet shape analysis can be performed by varying the parameter $Q_J$~\cite{Jouttenus:2013hs} in the definition of the final state jet momentum $P_J$ in Eq.~(\ref{pjet}). We see that increasing and decreasing $Q_J$ corresponds to including more and less radiation  as part of the final state jet momentum respectively. i.e. the value of $Q_J$ affects which particles are grouped into the jet region and which are grouped into the beam region. It also affects the geometric shape of the final state jet. Thus, varying $Q_J$ in the 1-jettiness framework is similar to choosing different jet algorithms or varying the jet radius in traditional jet algorithm based shape analyses. 

One can also define $Q_J$ in terms of the explicit jet algorithm used to find the jet reference vector $q_J$. For example, we can define~\cite{Jouttenus:2013hs} $Q_J$ as
\bea
Q_J &=& 2 \rho(R,y_K) K_{J_T} \cosh y_K,
\eea
where $K_{J_T}$ and $y_K$ are the transverse momentum and rapidity respectively, of the leading jet found by the explicit jet algorithm (see Eq.~(\ref{Kjet})) and the parameter $\rho(R,y_K)$ is a function of the jet radius $R$ used in the jet algorithm and $y_K$. All of the numerical results presented so far correspond to the choice $\rho(R,y_K)=1$. As explained in Ref.~\cite{Jouttenus:2013hs}, different choices for $\rho(R,y_K)$ yield different shapes for the jet in the $y-\phi$ plane. 

In order to illustrate the impact of varying $Q_J$, we consider the $\tau_1$-distribution for the choice
\bea
\rho(R=1.0,y_K) &=& f(y_K)= 0.834-0.233 y_K^2+0.077 y_K^4-0.008 y_K^6,
\eea
taken from one of the parameterizations in Ref.~\cite{Jouttenus:2013hs}. In Fig.~\ref{QJshape}, we plot the ratio
\bea
\label{RJ}
R_{Q_J} &\equiv&  \frac{d\sigma \big [\rho(R,y_K)=1\big ]_{NNLL+NLO}}{d\sigma \big [\rho(R=1.0,y_K) =  f(y_K)\big ]_{NNLL+NLO}},
\eea
corresponding to the ratio of the NNLL+NLO $\tau_1$-distributions for the different choices of $\rho(R,y_K)$. As before, we used $Q_e=90$ GeV and integrated over the range $[P_{J_T}^{\text{low}},P_{J_T}^{\text{high}}]=[20\>{\rm GeV}, 30\>{\rm GeV}]$ and $|y| < 2.5$, and integrated over $\tau_1$-bins of size 0.1
 GeV. Since we are only interested the effect of varying $Q_J$, we make the same scale choices for the two cross-sections in the ratio $R_{Q_J}$ in Eq.(\ref{RJ}),  and the error bars in Fig.~\ref{QJshape} arise from the scale variations in Eqs.~(\ref{muFO}) and (\ref{muSCET}).
 
We see in Fig.~\ref{QJshape}, that the ratio $R_J$, with the scale variation uncertainty included, deviates from unity. This shows that the $\tau_1$-distribution is sensitive to how radiation is clustered into the jet and beam regions, which can be controlled by varying $Q_J$. Thus, the 1-jettiness framework can be used to perform detailed studies of the final state QCD radiation, its clustering into jet and beam regions, the energy distribution or energy loss in the jet by studying $\tau_1$-distributions by varying $P_{J_T},y,$ and $Q_J$, and the effects of the nuclear medium by studying these distributions for a wide range of nuclear targets. We leave such detailed studies for future work.

\section{Conclusions}

The 1-jettiness event shape ($\tau_1$) for DIS processes, introduced in Refs.~\cite{Kang:2012zr,Kang:2013wca}, is a tool for precision studies of QCD and a probe of nuclear dynamics. It quantifies the pattern of final state radiation in single jet production in the DIS process $e^-  + N_A \to e^-  + J + X$, where $N_A$ denotes a nucleus with atomic weight $A$ and $J$ denotes the final state jet with transverse momentum $P_{J_T}$ and rapidity $y$.  The region $\tau_1\ll P_{J_T}$ corresponds to configurations with a single narrow jet and only soft radiation ($E\sim \tau_1$) between the beam and jet directions. This region requires a resummation of large Sudakov logarithm $\alpha^n\ln^{2m} (\tau_1/P_{J_T})$ for $m\leq n$. A factorization and resummation framework was developed for this purpose in Ref.~\cite{Kang:2012zr} and numerical results  at the NNLL level of accuracy have now been obtained~\cite{Kang:2013wca,Kang:2013nha}. The resummation region $\tau_1\ll P_{J_T}$ is insensitive to the type of jet algorithm used to find the leading jet which determines the jet reference vector used in the definition of $\tau_1$. This allows one to predict the $\tau_1$-spectrum in the resummation region without explicit use of any jet algorithm.

In this work, we extended previous results to include the NLO ($\sim \alpha_s$) prediction for the $\tau_1$-spectrum in the fixed-order region $\tau_1 \sim P_{J_T}$, along with its matching to the resummation region to provide the complete NNLL+NLO spectrum. The fixed-order region corresponds to configurations where hard radiation  is allowed between the beam and leading jet directions and can be sensitive to the specific jet algorithm used. Consequently, predicting the spectrum in the fixed-order region requires incorporating an explicit jet algorithm to find the leading jet. We provided numerical results for the NLO spectrum using the anti-$k_T$ jet algorithm with jet radius $R=1.0$. However, our code is flexible enough to use other types of jet algorithms. We performed several consistency checks including demonstrating the the cancellation of infrared divergences between the virtual and real emission contributions and a comparison with the SCET resummation result, expanded to the same order in perturbation theory. 

We made use of profile functions for the various renormalization scales appearing in the SCET resummation formula to smoothly match the resummation and fixed-order regions of the $\tau_1$-spectrum. We also incorporated a phenomenological model for the universal non-perturbative soft function in the region $\tau_1\sim \Lambda_{QCD}$. Putting these results together we have obtained the full NNLL+NLO $\tau_1$-spectrum. We also did a jet shape analysis, defined within the 1-jettiness framework, that allows one to study the effect of grouping different amounts of radiation in the finals state jet definition. Such an analysis can be used to study energy distributions and energy loss in jets and the effects of the nuclear medium on jet shapes. 

The analysis presented in this work can applied to data collected at HERA and in proposed future electron-ion colliders such as the EIC and the LHeC, as a tool for precision studies of QCD and a probe of nuclear dynamics.

\acknowledgements
We are grateful to Jianwei Qiu for many useful  discussions and comments during the course of this work and for reading over the manuscript. We also thank Daekyoung Kang, Chris Lee, and Iain Stewart for useful comments on the manuscript. SM also thanks the Erwin-Schr\"odinger-Institute for their hospitality during the ``Jets and Quantum Fields for LHC and Future Colliders (2013)"  program, where part of this work took place. This work was supported in part by the U.S. Department of Energy, Division of High Energy Physics, under contract DE-AC02-06CH11357 (XL) and the grants DE-FG02-95ER40896 (XL),  DE-FG02-08ER4153 (XL), and DE-AC52-06NA25396 (ZK) and by Northwestern University (SM).

\appendix

\section{Details of the NLO Calculation}
\label{secNLO}

Here we provide some more details  on the NLO calculation for reference. In particular, we list the results for the spin and color averaged squared matrix elements for the partonic channels in Eqs.~(\ref{LO}) and (\ref{NLO}). We also show the parameterization of the phase space used to isolate IR singularities, as discussed in section~\ref{nlocalc}.

We give formulae for the spin and color averaged squared matrix element of the leading order process
\bea
\label{LO1}
e^-(p_a) + q (p_b) \to e^-(p_e) + q (p_q),
\eea
one loop virtual corrections to this process, and the NLO real emission process 
\bea
\label{NLO1}
e^-(p_a) + q (p_b) \to e^-(p_e) + q (p_q) + g(k).
\eea
The result for the analogous process with an antiquark $\bar{q}$ is obtained by the replacement $p_q\to p_{\bar{q}}$. The result for the gluon initiated process
\bea
\label{NLOgluon}
e^- + g \to e^- + q + \bar{q},
\eea
is related to the process in Eq.(\ref{NLO1}) by crossing symmetry.

\subsection{LO Amplitude Squared}
The leading order spin and color averaged amplitude squared for the process in Eq.~(\ref{LO1}) is given by
\bea
|{\cal M}|_{0,d}^2 = 2 (4\pi \alpha )^2 Q_q^2 \frac{1}{\left(2p_a\cdot p_e\right)^2}\,
\left[
\left(2p_a\cdot p_b\right)^2 + \left(2p_q \cdot p_a\right)^2 - \epsilon \left(2p_a\cdot p_e\right)^2
\right],
\eea
in $d=4-2\epsilon$ dimensions.

\subsection{NLO Amplitude Squared}

The expansion in $\epsilon$ of the spin and color averaged  amplitude squared for the process in Eq.(\ref{NLO1}) is given by
\bea
\label{NLOa}
|{\cal M}|_{1,d}^2 &\equiv &  {\cal F}_{0 } + \epsilon {\cal F}_{1 } +  \epsilon^2 {\cal F}_{2 } + {\cal O} (\epsilon^3),
\eea
where the order $\epsilon^2$ term ${\cal F}_{2 }$ is given by
\bea
\label{F2}
{\cal F}_{2 }= 4g_s^2C_F (4\pi\alpha)^2 Q_q^2 \frac{1}{2p_a\cdot p_e} \,
\left(\frac{4(p_b\cdot k)^2+4(p_q \cdot k)^2}{4(p_b\cdot k)(p_q\cdot k )}-2 \right)\,,
\eea
 the  order $\epsilon$ term ${\cal F}_{1}$ is given by
\bea
\label{F1}
&&{\cal F}_{1}= (-8)g_s^2C_F (4\pi\alpha)^2 Q^2_q \frac{1}{2p_a\cdot p_e}\, 
\times \left\{\,
\frac{4(p_b\cdot k )^2+4(p_q\cdot k)^2+4(p_b\cdot p_q)(p_a\cdot p_e)}{4(p_b\cdot k)(p_q \cdot k)}\,
-1 \right\} \nn\\
+ &&
 (-8)g_s^2 C_F (4\pi\alpha)^2Q_q^2 \left(\frac{1}{2p_a\cdot p_e} \right)^2
(2p_a\cdot k)\times \left\{ 2\left(p_a\cdot p_b - p_a \cdot p_q \right) 
\left(\frac{1}{2p_q\cdot k }-\frac{1}{2p_b\cdot k}\right) \right. \nn \\
&&
\hspace{30.ex}
\left.
+ \frac{4(p_a\cdot k )(p_q\cdot p_b)}{4(p_b\cdot k)(p_q\cdot k)}
\right\}\,,
\eea
and the order  $\epsilon^0$ term ${\cal F}_0$ is given by
\bea
\label{F0}
{\cal F}_{0}=4g_s^2 C_F (4\pi\alpha)^2Q_q^2 \,
\frac{(2 p_a\cdot p_b)^2+(2p_a\cdot p_q)^2+(2p_e\cdot p_b)^2+(2p_e\cdot p_q)^2}{(2p_a\cdot p_e)(2p_b\cdot k)(2p_q \cdot k)} \, .
\eea

\subsection{Soft Limit}
In the limit that the final state gluon momentum $k$ in Eq.~(\ref{NLO1}) becomes soft, the spin and color averaged amplitude squared is given by the leading order squared amplitude multiplied by the eikonal factor
\bea
|{\cal M}|^2 = 2g_s^2C_F \frac{p_b\cdot p_q }{p_b\cdot k \>p_q\cdot k} |{\cal M}|^2_{0,d}.
\eea
One can check that this $k\to 0$ limit is correctly reproduced by Eqs.(\ref{NLOa}), (\ref{F2}), (\ref{F1}), and (\ref{F0}).

\subsection{Collinear Limit}

In the limit that the gluon is collinear to the incoming quark in Eq.(\ref{NLO1}), the spin and color averaged squared matrix element is given by the leading order squared matrix element multiplied by 
the splitting kernel
\bea
|{\cal M}|^2 = 2g_s^2C_F \frac{1}{2p_g\cdot p_b(1-z)}P_{gq}(z,\epsilon) \,
|{\cal M}|^2_{0,d}\,,
\eea
with $p_b$ in $|{\cal M}|^2_{0,d}$ being replaced by $p_b' = (1-z)p_b$, where $z = E_g/E^q_b$ and
\bea
P_{gq}(z,\epsilon) = \frac{1+(1-z)^2}{z} - \epsilon z .
\eea 
Similarly, for the case where the gluon is collinear with the outgoing quark in Eq.(\ref{NLO1}), we have
\bea
|{\cal M}|^2 = 2g_s^2C_F \frac{1}{2p_g\cdot p_q}P_{gq}(z,\epsilon) \,
|{\cal M}|^2_{0,d}\,,
\eea
with $p_q$ in $|{\cal M}|^2_{0,d}$ replaced by $p_q' = p_q/(1-z)$ and
$z = E_g/(E_g + E_q)$. One can explicitly check that these collinear limits are correctly reproduced from  the full squared amplitude in Eqs.(\ref{NLOa}), (\ref{F2}), (\ref{F1}), and (\ref{F0}).

\subsection{Virtual NLO correction}
The NLO virtual correction to the squared amplitude for the leading order process in Eq.(\ref{LO1}) is given by
\bea
\label{virtualpole}
\left(\frac{e^{\gamma_E}\mu^2}{4\pi}\right)^\epsilon |{\cal M}_V|^2 = |{\cal M}|^2_{0,d} \left[\,
1+\frac{\alpha_s}{2\pi}C_F\left(\,
-\frac{2}{\epsilon^2} - \frac{3}{\epsilon} - \frac{2}{\epsilon}L\,
-L^2 - 3L - 8 + \frac{\pi^2}{6}
\right)
\right]\,,
\eea
after UV renormalization so that the poles correspond to IR diveregences and we have used the notation $L = \log \frac{\mu^2}{Q^2}$, where $Q^2=-(p_a-p_e)^2$.

\subsection{Gluon Initiated Channel}

The spin and color averaged amplitude squared for the gluon initiated channel in Eq.(\ref{NLOgluon}) can be obtained from the results for the channel in Eq.(\ref{NLO1}) by crossing symmetry
\bea
|{\cal M}|^2_{eg} = |{\cal M}|^2_{eq}(\cdots,p_b, k ,\cdots) \mapsto \,
 -\frac{1}{1-\epsilon}\frac{N_c}{N_c^2-1}|{\cal M}|^2_{eq}(\cdots,- k, -p_b ,\cdots) \,.
\eea
The factor $N_c/(N_c^2-1)$ is needed to convert the color-averaging factor  $1/N_c$ for an initial state quark to $1/(N_c^2-1)$ for  an initial state gluon. Similarly, the factor of $1/(1-\epsilon)$ is needed to convert the spin-averaging factor $1/2$ for an initial state quark to $1/(d-2)=1/(2-2\epsilon)$ for an initial state gluon.
In this channel, there are only collinear IR divergences arising when either the final state quark or anti-quark becomes collinear with the initial gluon. In this channel,  we only have single poles since there are only collinear divergences, so that we only need to keep the matrix element squared up to ${\cal O}(\epsilon)$. 


\section{Phase Space parameterization and Isolating IR Singularties}
\label{psdetail}

The partonic cross-section is given by the spin and color averaged cross-section 
\bea
\hat{d\sigma} \equiv \int dPS \> \hat{{\cal F}}_{\text{meas.}}([PS])\> |\overline{{\cal M}}|^2,
\eea
where $\hat{{\cal F}}_{\text{meas.}}$ denotes the measurement function that acts of the final state phase space. Note that the phase space measure $dPS$ and the spin and color average amplitude squared $|\overline{{\cal M}}|^2$ are both Lorentz invariant quantities. However, in general the measurement function  $\hat{{\cal F}}_{\text{meas.}}([PS])$ is not Lorentz invariant. In the following, we first parameterize the phase space measure $dPS$ and the momenta and scalar products in $|\overline{{\cal M}}|^2$ in the partonic center of mass frame. In the final step, just before imposing the final state restrictions through the measurement function $\hat{{\cal F}}_{\text{meas.}}([PS])$ and performing the phase space integrations, we perform a boost  to the hadronic center of mass frame, since the restrictions in  $\hat{{\cal F}}_{\text{meas.}}$ are in terms of these boosted momenta. 

In the partonic center of mass frame, the initial state momenta are given by
\bea
p_a = \frac{E_{\rm cm}}{2}(1,0,0,-1)\,, \hspace{5.ex}
p_b = \frac{E_{\rm cm}}{2}(1,0,0,1)\,, \hspace{5.ex}
\eea
where $p_a$ and $p_b$ denote the momentum of the initial electron and parton respectively and $E_{\rm cm}$ denotes the partonic center of mass energy. In this convention, the initial quark momentum is along the positive z-axis direction. 

For the leading order process in Eq.(\ref{LO1}), the final state phase space is given by
\bea
\int \mathrm{d}{\rm PS} = \frac{1}{2\hat{s}}\int \frac{\mathrm{d}^{d-1}p_q}{2(2\pi)^{d-1}E_q}\,
\frac{\mathrm{d}^dp_e}{(2\pi)^{d-1}}\delta^+(p_e^2)\,
(2\pi)^{(d)}(p_a+p_b - p_e - p_q) \,,
\eea
where $\hat{s}=E_{\rm cm}^2$.  The electron phase space integration can be performed using the four-momentum conserving delta function and the remaining delta function $\delta^+(p_e^2)$ can
be removed by integrating over the magnitude of the quark momentum $|p_q|=E_q$, which gives
\bea
\label{PS-a}
\int \mathrm{d}{\rm PS} = \frac{1}{2\hat{s}}\,
\int \frac{\mathrm{d}c_q\, \mathrm{d}\Omega^q_{d-2}}{8(2\pi)^{d-2}}\,
\left(\frac{E_{\rm cm}}{2}s_q\right)^{-2\epsilon}\,
\equiv \frac{1}{2s} \int \mathrm{d} {\rm Lips}_{ab\to p_e + p_q}\,
\left(p^T_q\right)^{-2\epsilon}\,
\,,
\eea
where $c_q = \cos \theta_q, s_q=\sin \theta_q$, $p_q^T \equiv \frac{E_{\rm cm}}{2}s_q$, and $\theta_q$ is the angle between the final state quark and the positive z-axis direction. The unit three vector ${\bf n}_q$ along the final state quark momentum ${\bf p}_q$ is given by ${\bf n}_q = (s_q \cos\phi_q,s_q \sin\phi_q,c_q)$, where $\phi_q$ is the quark azimuthal angle around the z-axis.  

Usually the integrand of the phase space integration is independent of the azimuthal angle $\phi_q$ and this is true for our case as well. We choose $\phi_q = 0 $ and integrate over $\Omega^q_{d-2} $. As explained below, we can then reintroduce the azimuthal angle $\phi_q$ to facilitate comparison with the NLO case where an extra parton is emitted in the final state.
We parameterize the angular integration over $\theta_q$ as $c_q = 1-2x_{q,1}$, where the new integration variable $x_{q,1}$ has the range of integration $[0,1]$. One can then perform the integral over $d\Omega^q_{d-2}$, which includes the azimuthal integration over $\phi_q$, to get
\bea
\mathrm{d} {\rm Lips}_{ab\to p_e + p_q} \to \frac{(4\pi)^{\epsilon}}{\Gamma(1-\epsilon)}\,
\mathrm{d} {\rm Lips}_{ab\to p_e + p_q} = \,
\frac{(4\pi)^{\epsilon}}{\Gamma(1-\epsilon)}\,
\frac{\mathrm{d}x_{q,1}}{8\pi}\,.
\eea
We can now reintroduce the azimuthal integration by defining $\phi_q = 2\pi x_{q,2}$ so that $x_{q,2}$ has the range of integration $[0,1]$ and we can write
\bea
\mathrm{d} {\rm Lips}_{ab\to p_e + p_q} \to   \,
\frac{(4\pi)^{\epsilon}}{\Gamma(1-\epsilon)}\,
\frac{\mathrm{d}x_{q,1}\,\mathrm{d}x_{q,1}}{8\pi}\,.
\eea

For the NLO  real emission process $p_a +p_b \to p_e + p_q + k$, the final state phase space  integral has the form
\bea
\int \mathrm{d}{\rm PS} = \frac{1}{2\hat{s}}\int 
\frac{\mathrm{d}^{d-1}k}{2(2\pi)^{d-1}E_g}\,
\frac{\mathrm{d}^{d-1}p_q}{2(2\pi)^{d-1}E_q}\,
\frac{\mathrm{d}^dp_e}{(2\pi)^{d-1}}\delta^+(p_e^2)\,
(2\pi)^{(d)}(p_a+p_b - p_e - p_q - k) \,. \nn \\
\eea
Once again, integrating over the electron phase space using the four momentum conserving delta function and over the quark energy $|p_q|=E_q$ using $\delta^+(p_e^2)$, we get  \bea
\label{PSg}
\int \mathrm{d}{\rm PS} = \frac{1}{2\hat{s}}\int 
\frac{\mathrm{d}^{d-1}k}{2(2\pi)^{d-1}E_g}\,
\frac{\mathrm{d}x_{q,1}\, \mathrm{d}x_{q,2}}{8\pi}\,
\frac{2E_q}{(Q^0 - {\bf n}_q\cdot {\bf Q})}\left(\frac{E_qs_q}{p^T_q}\right)^{-2\epsilon}\,
\frac{(4\pi)^{\epsilon}}{\Gamma(1-\epsilon)}\,
\left(p^T_q\right)^{-2\epsilon}\,
\,,
\eea
where we have defined 
\bea
Q \equiv p_a+p_b-k \, .
\eea 
The final state quark energy in terms of $Q$ is given by
\bea
E_q = \frac{Q^2}{2(Q^0 - {\bf n}_q\cdot {\bf Q})}\,,
\eea
and the $x_{q,i}$, $p^T_q$ and ${\bf n}_q$ have the same definitions as in the case of the leading order process.

The IR singularities arise when the final state gluon is soft ($E_g\to 0$) or collinear with either $p_b$ or $p_q$.  
Following the sector decomposition technique~\cite{Frixione:1995ms,Czakon:2010td,Boughezal:2011jf,Boughezal:2013uia}, we introduce the partition
\bea
\Delta^{g j }_\theta = \frac{(n_g \cdot n_i)^\alpha}{(n_g\cdot n_b)^\alpha + (n_g \cdot n_q)^\alpha}\,,
\eea
with $\alpha \ge 1$, $j\ne i$,  and $i$, $j = b$, $q$ to isolate the two different collinear singularites. We choose $\alpha=1$. We note that
\bea
\Delta^{g b}_\theta + \Delta^{g q}_\theta = 1\,,
\eea
and the phase space can be partitioned into two sectors
\bea
\label{PSsector}
d{\rm PS} = d{\rm PS}^{(gb)} + d{\rm PS}^{(gq)}\,,
\eea
where each sector is defined as
\bea
\label{twosectors}
\mathrm{d}{\rm PS}^{(gb)} \equiv  \mathrm{d}{\rm PS}\>\Delta^{g b}_\theta, \qquad \mathrm{d}{\rm PS}^{(gq)}\equiv d{\rm PS} \>\Delta^{g q}_\theta \, .
\eea
In the $\mathrm{d}{\rm PS}^{(gb)}$ sector, the collinear singularity arising when the final state gluon and quark become unresolved is canceled by the partition factor $\Delta^{g b}_\theta$. Similarly, in the $\mathrm{d}{\rm PS}^{(gq)}$ sector, the collinear singularity arising when the final state gluon and the initial state quark become unresolved is  canceled by $\Delta^{g q}_\theta$. These two types of collinear singularities occur in different regions of phase space. The sector decomposition ensures that only one of these collinear singularities occurs in each sector. This allows one to adapt a convenient parameterization of the phase space in each sector to isolate the corresponding collinear singularity in that sector.

For the $\mathrm{d}{\rm PS}^{(gb)} $ sector, we parametrize the final state gluon momentum $k$ as
\bea
\label{glumom}
k = x_{g,1}\frac{E_{\rm cm}}{2}(1, s_g\cos \phi_g,s_g \sin \phi_g,c_g)\,,
\eea
where $c_g =\cos \theta_g, s_g=\sin \theta_g$ and the angles $\theta_g,\phi_g$ are defined relative to the positive z-axis direction as before. The relative azimuthal angle between the final state gluon and quark is denoted by $\tilde{\phi}_g = \phi_g  -\phi_q $. Using this parameterization, the gluon phase space measure can be brought into the form
\bea
[\mathrm{d}g]^{(gb)} \equiv \frac{\mathrm{d}^{d-1}k}{2(2\pi)^{d-1}E_g} \Delta^{g b}_\theta= \frac{\mathrm{d}x_{g,1}\mathrm{d}c_g \mathrm{d}{\phi}_g\,
\Omega^g_{d-3}}{2(2\pi)^{d-1}}\left(s_g^2 \sin^2 \tilde{\phi}_g\right)^{-\epsilon} x_{g,1}^{1-2\epsilon}\,
\left(\frac{E_{\rm cm}}{2}\right)^{2-2\epsilon} \,\Delta^{g b}_\theta \,  . \nn \\
\eea
Introducing the $x_{g,2}, x_{g,3}$ variables defined by $c_g = 1- 2x_{g,2}$ and $\phi_g = 2\pi x_{g,3}$ respectively and with a range of integration $[0,1]$, the gluon phase space measure becomes
\bea
\label{gb}
[\mathrm{d}g]^{(gb)} = 
\left(\frac{E_{\rm cm}}{2}\right)^{2-2\epsilon}\,
\frac{2^{-2\epsilon}\,
\Omega^g_{d-3}}{(2\pi)^{d-2}}\left( \sin^2 \tilde{\phi}_g\right)^{-\epsilon} x_{g,1}^{1-2\epsilon}\,
x_{g,2}^{-\epsilon}(1-x_{g,2})^{-\epsilon}
 \,\Delta^{g b}_\theta \,
\prod_{i=1}^3 \, \mathrm{d} x_{g,i}
\,.
\eea
The solid angle factor $\Omega^g_{d-3}$ can be expanded in $\epsilon$ to give
\bea
\label{solid}
\Omega^g_{d-3} = \frac{2\pi^{1-\epsilon}}{\Gamma(1-\epsilon)}\,
 \left( \frac{2\pi^{1/2}\Gamma(1/2-\epsilon)}{\Gamma(1-\epsilon)}\right)^{-1}\,
= \left(1-\frac{\pi^2}{3}\epsilon^2\right) 2^{-2\epsilon}\pi^{-\epsilon}
 \Gamma(1+\epsilon) + {\cal O}(\epsilon^3) \,.
\eea
Using Eqs.(\ref{gb}) and (\ref{solid}), the total phase space measure for the $[gb]$ sector as defined in Eqs. (\ref{PSg}), (\ref{PSsector}), and (\ref{twosectors}) is given by
\bea
\mathrm{d}{\rm PS}^{(gb)} = \frac{1}{2\hat{s}} \mathrm{d}{\rm Lips}^{(gb)}\,,
\eea
where
\bea\label{PS}
\mathrm{d}{\rm Lips}^{(gb)} = {\rm Norm}\times {\rm PS}^{(gb)}_{w} \,
\times \left({\rm ePS}^{(gb)}\right)^{-\epsilon}\,
\times \prod_{i=1}^{3} \mathrm{d}x_{g,i} \prod_{i=1}^2 \mathrm{d}x_{q,i} \,
x_{g,1}^{-1-2\epsilon}\,
x_{g,2}^{-1-\epsilon}\,
\times \left[x_{g,1}^2 x_{g,2} \right] \, .\nn \\
\eea
The factors ${\rm Norm},{\rm PS}^{(gb)}_w,$ and $ {\rm ePS}^{(gb)} $ are given by
\bea
&&{\rm Norm} = \frac{\Gamma(1+\epsilon)}{16\pi^2(4\pi)^{-\epsilon}}
\left( 1- \frac{\pi^2}{3}\epsilon^2\right) \times
\frac{(4\pi)^{\epsilon}}{\Gamma(1-\epsilon)}\,
\left(p^T_q\right)^{-2\epsilon}\,
\,, \nn \\
&&{\rm PS}^{(gb)}_w = \,
\frac{1}{8\pi}\frac{2 E_{\rm cm}^{2} E_q}{(E_{\rm cm} - E_g(1 - {\bf n}_q \cdot {\bf n}_g))}\,
 \Delta^{g b}_\theta \,,\nn \\
&& {\rm ePS}^{(gb)} = \,
4 E_{\rm cm}^2 \sin^2(\phi_g - \phi_q) (1-x_{g,2}) \left(\frac{E_q s_q}{p_q^T}\right)^2\,.
\eea
All the final state parton momenta can now be generated in terms of specified values for $E_{\rm cm}$ and the integration variables $x_{g,i},x_{q,i}$. 

\OMIT{We note that
if we set the combination $\frac{(4\pi)^{\epsilon}}{\Gamma(1-\epsilon)}
\left(p^T_q\right)^{-2\epsilon}$ to $1$ in both leading order phase space and the real emission phase space, the finite results
will not change.}

For the $d{\rm PS}^{(gq)}$ sector, we can choose the z-axis to align with $p_q$ and use the same parametrization as before. i.e. in the $d{\rm PS}^{(gb)}$ sector, the final state gluon momentum was parameterized with the initial parton along z-axis. By now setting the z-axis along $p_q$, the same parameterization as in Eq.~(\ref{glumom}) can be used for the $d{\rm PS}^{(gq)}$ sector. Once this is done, we can rotate back to the partonic center of mass frame. The net effect of this procedure is that the parameterization of $d{\rm PS}^{(gq)}$ is the same as that of ${\rm PS}^{(gb)}$ but with the replacement $\sin^2(\phi_g - \phi_q) \to \sin^2(\phi_g )$.  This can be understood as follows. Choosing the z-axis to align with $p_q$ so that the direction of $p_q$ is along the vector $\tilde{n}_q = (1,0,0,1)$, the parameterization of the gluon momentum $\tilde{k}$ in this frame  will take the same form as in the ${\rm dPS}^{(gb)}$ sector
\bea
\tilde{k} = x_{g,1}\frac{E_{\rm cm}}{2}(1, s_g\cos \phi_g,s_g \sin \phi_g,c_g)\,,
\eea
with $c_g = 1 - 2x_{g,2} $ and $\phi_g = 2\pi x_{g,3}$. 
\OMIT{
\bea
\tilde{k} = E_g(1,s_g \cos(\phi_g), s_g \sin(\phi_g), c_g)\,,
\eea}
The parameterization of the gluon momentum in ${\rm dPS}^{(gq)}$ will then be identical to that of ${\rm dPS}^{(gb)}$ but with the replacement
$\sin^2(\phi_g - \phi_q) \to \sin^2(\phi_g )$, since in this frame with the z-axis aligned with the quark momentum, the azimuthal angle of the gluon is directly measured relative to the final state quark momentum $p_q$. The gluon momentum $k$ in the center of mass frame is obtained via a rotation matrix defined by
\bea
{\bf n}_q = R \cdot \tilde{\bf n}_q,
\eea
where  ${\bf n}_q = (s_q \cos\phi_q,s_q \sin\phi_q,c_q)$ as before, so that
\bea
{\bf k} = R(x_{q,1},x_{q,2}) \tilde{\bf k} \,.
\eea
Since ${\rm det}(R) = 1$ and the phase space measure is rotationally invariant, the same result of the ${\rm dPS}^{(gb)}$ sector will be reproduced with the replacement
$\sin^2(\phi_g - \phi_q) \to \sin^2(\phi_g )$.


\bibliographystyle{h-physrev3.bst}
\bibliography{nlodis}

\end{document}